\documentclass[aps,prd,reprint,groupedaddress,floatfix]{revtex4-1}

\usepackage{graphics} 
\usepackage{amsbsy}

\begin{document}


\title{QED radiative effects in the processes of exclusive photon electroproduction from polarized protons with the next-to-leading accuracy}


\author{Igor Akushevich}
\email[]{igor.akushevich@duke.edu}
\affiliation{Physics Department, Duke University, Durham, NC 27708, USA\\ and \\ Jefferson Lab., Newport News, VA 23606, USA}
\author{Alexander Ilyichev}
\affiliation{National Center for Particle and High Energy Physics, Byelorussian State University, Minsk, 220088, Belarus}
\author{Nikolai M. Shumeiko}
\affiliation{National Center for Particle and High Energy Physics, Byelorussian State University, Minsk, 220088, Belarus}

\date{\today}

\begin{abstract}
Radiative effects in the electroproduction of photons in  polarized $ep$-scattering are calculated with the next-to-leading (NLO) accuracy. The contributions of loops and two photon emission were presented in analytical form. The covariant approach of Bardin and Shumeiko was used to extract the infrared divergence. All contributions to the radiative correction were presented in the form of the correction to the leptonic tensor thus allowing for further applications in other experiments, e.g., deep inelastic scattering. The radiative corrections (RC) to the cross sections and polarization asymmetries were analyzed numerically for kinematical conditions of the current measurement at Jefferson Lab. Specific attention was paid on analyzing kinematical conditions for the process with large radiative effect when  momenta of two photons in the final state are collinear to momenta of initial and final electrons, respectively.  
\end{abstract}


\maketitle
\section{\label{Intro}Introduction}

\begin{figure}[b]\centering
\scalebox{0.32}{\includegraphics{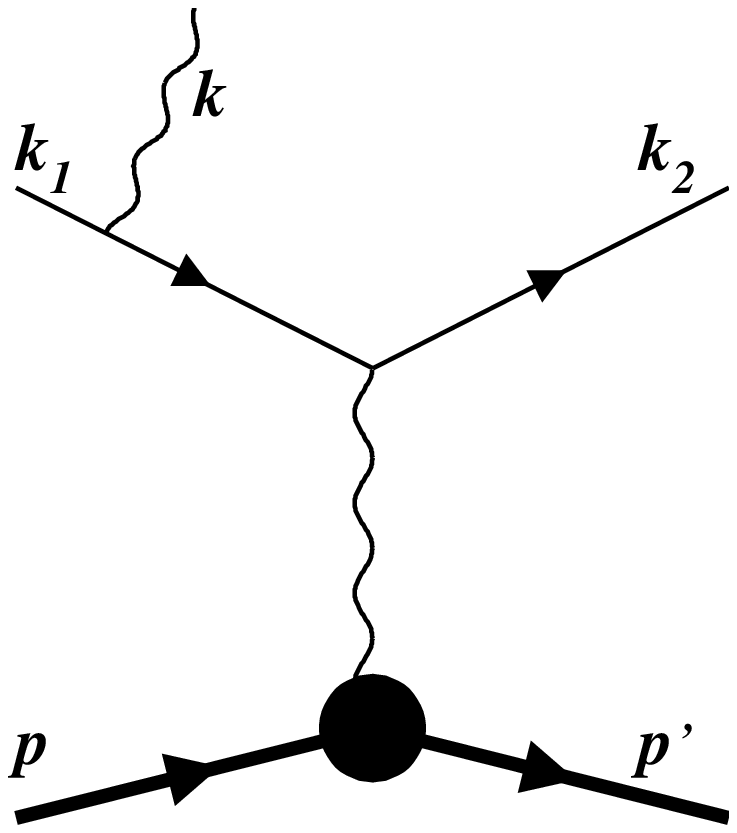}}
\hspace{0.4cm}
\scalebox{0.32}{\includegraphics{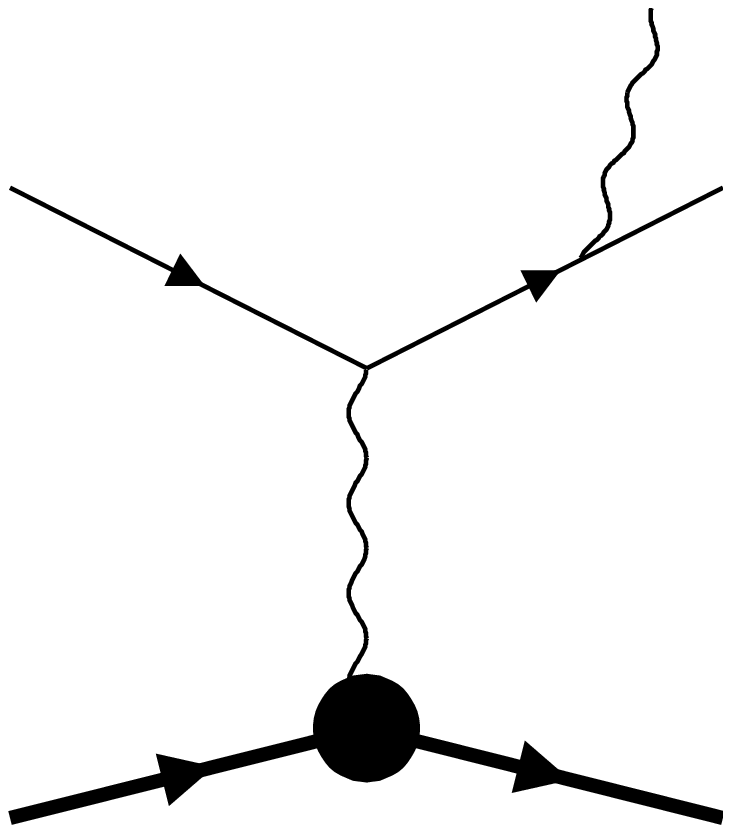}}
\\
{\bf a) \hspace{2cm} b)}
\caption{\label{BHgraphs}Feynman graphs of BH process}
 \end{figure}

The experimental precision attainable in modern experiments on $ep$-scattering allows for measuring the cross section and spin asymmetries in the Bethe-Heitler (BH) process with an unprecedented accuracy. The QED radiative correction is one serious source of systematical uncertainties and therefore must be known with  any predetermined accuracy. Several available calculations provide only leading logarithm accuracy and/or have certain other limitations. The most comprehensive calculation was presented by Vanderhaeghen with colleagues \cite{Vanderhaeghen2000}. One feature of the calculation is the detailed consideration of the one-loop correction. The basic limitations include i) only approximated evaluation of the radiative tail corresponding to additional photon emission processes, where the photon energy is very small compared to the lepton momenta and ii) not fully analytical calculation of loop effects. Bytev et al. \cite{ByKuTo2008PRC} used the electron structure functions for calculation of the hard 
photon emission in the leading approximation. The authors described the approach for the calculation in the leading log approximation in necessary details and gave useful expressions for the kinematical variables in so-called ``shifted'' kinematics. No explicit formulas were presented for the standard measurements of the the BH process but for the cross section integrated over the energy fraction of scattered electron. The calculation of our group \cite{AkushevichIlyichev2012} was also performed in the leading log approximation. The quality of this approximation is difficult to control without the next-to-leading contribution or exact calculations.   
 
The structure of the dependence of the RC cross section on the electron mass is 
$
\sigma_{RC}=A\log({Q^2}/{m^2})+B+O(m^2/Q^2),
$
where $A$ and $B$ do not depend on the electron mass. For experiments with transferring momentum squared of one GeV squared ($Q^2 \sim$ 1GeV$^2$), $\log\bigl({Q^2}/{m^2}\bigr) \sim 15$. In practice the leading accuracy can be not so good because often $ B>A$ or even $ B\gg A$, especially for the contribution with additional photon emission. In this paper we calculate RC with the next-to-leading accuracy. In this approximation both $A$ and $B$ are calculated exactly. The terms neglected in this approximation are of order of the electron mass squared and therefore the  calculation of the lowest order RC to BH cross sections and spin asymmetries presented in this paper is practically exact. Note that neglecting the terms with the electron mass is not fundamental, and the completely exact approach of the RC to the BH process within the NLO accuracy can be reconstructed in the case of need.

Methodologically, our calculation is based on two sources. The first is the calculation of the QED RC to radiative tail from elastic peak presented in \cite{AkBaSh1986YP}. We largely follow the methods suggested in this paper. The second is the results of analytical integration for one loop contribution given in \cite{Kuraevintegrals}. We use the analytical form for results of integration over a loop momentum suggested in this paper.    

The RC calculated in a certain kinematical point depend on the way on how the kinematical variables representing the kinematical point were reconstructed. For example, the virtual photon momentum squared (i.e., $-t$) can be reconstructed using measured momentum of the final proton or, alternatively, using the momenta of the final electron and photon. Radiation of an additional unobserved photon from the lepton line results in the same reconstructed value of $t$ in the first case and in change of this value in the second case. Therefore, the effect of RC is expected to be different in these two cases. The results presented in this paper are valid for the case when the kinematics of the BH event is reconstructed using final momenta of the final electron and proton only. However, this restriction is not critical because the results for another scheme of reconstruction of kinematical variables can be reproduced using a Monte Carlo program. In such a program the results obtained for a certain scheme of 
reconstruction of kinematical variables (e.g., the results obtained in this paper) can be used to generate  events with and without an additional radiated photon and then another reconstruction scheme is applied for the generated sample of events resulting in estimating RC effects for this reconstruction scheme. 

The remainder of the paper is organized as follows. Exact and approximate expressions for the BH cross section are presented in
Section \ref{sectioncrosssection}. The one-loop contributions to the total BH cross section are discussed in section \ref{Loops}. The cross section of emission of the two photon 
is presented in Section \ref{RC}. Specific attention is paid to critical components of the approach including infrared divergence extraction and reproduction of the known formulas in the leading log approximation. 
All contributions are combined in section \ref{ocs123} resulting in explicit expressions for the radiative-corrected  BH cross section. Detailed numerical analysis 
is given in Section \ref{na321} focusing on analysis of RC in the current experiments with unpolarized and polarized particles at JLab and on revealing the kinematical regions where the RC is large. Discussion and conclusions 
are presented in Section \ref{DC}. The paper includes four Appendices: the table of loop integrals is calculated in Appendix \ref{loopintegrals}; the most cumbersome explicit results for the loop contributions are given in Appendix \ref{Vresults}; 
integrals over the angles of two real photons are presented in Appendix \ref{Rintegrals}; and explicit results of the two-photon bremsstrahlung cross section are given in Appendix \ref{Rappendix}.

\section{\label{sectioncrosssection}The lowest order BH cross section}

The BH process 
\begin{equation}\label{BHprocess}
e(k_1,\xi)+p(p,\eta)\longrightarrow e'(k_2)+p'(p')+\gamma(k),
\end{equation}
is traditionally described using four kinematical variables: $Q^2=-(k_1-k_2)^2$, $x=Q^2/(2p(k_1-k_2))$, $t=-(p-p')^2$, and $\phi$, the angle between planes 
$({\bf k_1},{\bf k_2})$ and $({\bf k_1}-\bf{k_2},{\bf p'})$. Polarizations of the initial electron and proton are described by 4-vectors $\xi$ and $\eta$. The cross section of the BH process for unpolarized and polarized targets is
\begin{eqnarray}\label{dsigma0}
d\sigma_0&=&\frac 1{2\sqrt{\lambda_S}} {\cal M}_{BH}^2 d\Gamma_0.
\end{eqnarray}
We use the phase space representation: 
\begin{eqnarray}\label{dGamma0}
d\Gamma_0&=&\frac 1{(2\pi)^5}
\frac{d^3k_2}{2E_2}
\frac{d^3p'}{2p'_0}
\frac{d^3k}{2\omega}
\delta^4(k_1+p-k_2-p'-k)
\nonumber \\&=&
\frac{Q^2dQ^2dxdtd\phi}{(4\pi)^4x^2 \sqrt{\lambda_S} \sqrt{\lambda _Y}}.
\end{eqnarray}
 Here $S=2ME_1$, $m$ and $M$ are the lepton and proton masses, $E_1$ is the beam energy, and kinematical invariants are defines as 
\begin{eqnarray}
&&S_x=S-X=Q^2/x, \quad S_p=S+X, \quad S_{xt}=S_x-t.
\nonumber\\
&&\lambda_S=S^2-4m^2M^2, \quad \lambda _Y=S_x^2+4 M^2Q^2. 
\end{eqnarray}
The $\phi$-dependence appears only through its occurrence in invariants $w_0=2kk_1$ and $u_0=2kk_2$:
\begin{eqnarray}\label{uwexplicit00}
w_0&=&\frac{1}{2}(t-Q^2)+{S_p\over 2\lambda_Y}\bigl(S_x(Q^2+t)
-2tQ^2\bigr)
\nonumber\\&&\quad
+{\sqrt{\lambda_{uw}} \over \lambda_Y} \cos\phi
\end{eqnarray}
and $u_0=w_0+Q^2-t$ (it explicitly results in the same expression as for $w_0$ but with the opposite sign at the first term). Also,
\begin{eqnarray}\label{lambdav00}
\lambda_{uw}&=&4(SXQ^2-M^2Q^4-m^2\lambda_Y)
\nonumber\\&&
\times\bigl(tS_{xt}(S_x-Q^2)
-M^2(Q^2-t)^2\bigr).
\end{eqnarray}
Alternatively, the azimuthal angle of the photon $\phi_k$
 can be used instead of $\phi$. In this case $\cos\phi=-\cos\phi_k$.

The BH matrix element is 
${\cal M}_{BH}=e^3t^{-1}J^h_\mu J_{\mu}^{BH}$ with 
\begin{equation}
J^h_\mu={\bar u}(p')\biggl(\gamma_{\mu}F_1+i\sigma_{\mu\nu}\frac{p_\nu '-p_\nu}{2M}F_2\biggr)u(p)
\end{equation}
and
\begin{eqnarray}
J_{\mu}^{BH}&=& 
{\bar u}_2\Biggl [
 \gamma_\mu \frac{{\hat k}_1-{\hat k}+m}{-2kk_1}{\hat \epsilon}
+ {\hat \epsilon} \frac{{\hat k}_2+{\hat k}+m}{2kk_2}\gamma_\mu 
\Biggr ]u_1
\nonumber \\&=& -
{\bar u}_2\Biggl [\left(\frac {k_1\epsilon}{kk_1}-\frac {k_2\epsilon}{kk_2}\right)\gamma_\mu
-\frac{\gamma_\mu \hat{k}\hat{\epsilon}}{2kk_1}
-\frac{\hat{\epsilon}\hat{k}\gamma_\mu }{2kk_2}
\Biggr ]u_1, 
\end{eqnarray}
where ${\bar u}_2\equiv {\bar u}(k_2)$, ${u}_1\equiv {u}(k_1)$,  $\epsilon$ is the photon polarization vector, and $F_{1,2}$ are the proton formfactors.


The key intermediate quantity in this paper is the leptonic tensor. The representation of the RC at the level of the leptonic tensor is one convenient way of having compact analytical expressions for all contributions to RC. The BH cross section,
\begin{eqnarray}
d\sigma_0&=&\frac{32\pi^3\alpha^3}{\sqrt{\lambda_S}t^2}
 \bigl(J^h_\mu J^{BH}_{\mu}\bigr)^2 d\Gamma_0,
\end{eqnarray}
can also be represented in terms of convolution of the leptonic and hadronic tensors, 
\begin{eqnarray}
d\sigma_0&=&\frac{32\pi^3\alpha^3}{\sqrt{\lambda_S}t^2}
 L_{\mu\nu}W_{\mu\nu} d\Gamma_0.
\end{eqnarray}
The hadronic tensor is defined as 
$W_{\mu\nu}={\mathrm Tr} \sum J^h_\mu J^{h\dagger} _\nu$ 
where the symbol of sum means averaging and summing over polarization states of the initial and final protons, respectively. The tensor is expressed as the sum of unpolarized (spin-averaged) and polarized (spin-dependent) parts:
\begin{equation}\label{wmunu}
W_{\mu\nu}=W_{\mu\nu}^{unp}+iW_{\mu\nu}^{pol}.
\end{equation}
Straightforward calculation results in
\begin{eqnarray}\label{wmunuunp}
W_{\mu\nu}^{unp}&=&
-t (F_1+F_2)^2 \tilde{g}_{\mu\nu}+4(F_1^2+\tau F_2^2)\tilde{p}_\mu \tilde{p}_\nu,
\end{eqnarray}
where $\tau=t/4M^2$ and
\begin{equation}
\tilde{g}_{\mu\nu}={g}_{\mu\nu}-{q_\mu q_\nu \over q^2}, \quad
\tilde{a}_{\mu}={a}_{\mu}-q_\mu {aq\over q^2} \quad
\end{equation}
for any four-vector $a$. Here and below $q=p'-p=k_1-k_2-k$.
The spin-dependent part of hadronic tensor is
\begin{eqnarray}\label{wmunupol}
W_{\mu\nu}^{pol}&=&2M(F_1-\tau F_2)(F_1+F_2)\epsilon_{\mu\nu \eta q}
\nonumber\\&&
+(F_1+F_2)F_2 {\eta q \over M}\epsilon_{\mu\nu p q}.
\end{eqnarray}
The leptonic tensor ($L_{\mu\nu}^{BH}={\mathrm Tr} \sum J^{BH}_\mu J^{BH \dagger}_\nu$) is also represented as a sum of unpolarized and polarized parts:
\begin{equation}
L_{\mu\nu}^{BH}=L_{\mu\nu}^{unp}+iL_{\mu\nu}^{pol}.
\end{equation}
Straightforward calculation of the unpolarized part of the leptonic tensor results in  
%
\begin{eqnarray}\label{leptonicunp}
L^{unp}_{\mu\nu}&=&\tilde{g}_{\mu\nu}T_1^0
+\tilde{k}_\mu \tilde{k}_\nu T_2^0
+\tilde{n}_\mu \tilde{n}_\nu T_3^0
\nonumber\\&&
+(\tilde{k}_\mu \tilde{n}_\nu+\tilde{n}_\mu \tilde{k}_\nu) T_4^0,
\end{eqnarray}
where $n=k_1+k_2$. If the calculation is conducted with the next-to-leading accuracy, the tensor is reduced to
\begin{eqnarray}
L^{unp}_{\mu\nu}&=&{2\over u_0w_0} \Bigl( -\tilde{g}_{\mu\nu}(2tQ^2+w_0^2+u_0^2)
\nonumber\\&&\qquad
+2(\tilde{k}_\mu \tilde{k}_\nu +\tilde{n}_\mu \tilde{n}_\nu ) t\Bigr).
\end{eqnarray}
Thus, $T^0_i$ in the NLO approximation are expressed as:
\begin{eqnarray}
&& T^0_1=-2\biggl(\frac{u_0}{w_0}+\frac{w_0}{u_0}\biggr)-\frac{4tQ^2}{u_0w_0}, \nonumber \\
&& T^0_2=T^0_3=\frac{4t}{u_0w_0}, \qquad T^0_4=0.  
\end{eqnarray}
Spin-dependent part of the leptonic tensor is
\begin{eqnarray}\label{leptonicpol0}
L^{pol}_{\mu\nu}&=&
\epsilon_{\mu \nu k q}T^{p0}_1
+\epsilon_{\mu\nu n q}T^{p0}_2
+\epsilon_{\mu\nu q \xi}T^{p0}_3,
\end{eqnarray}
where
\begin{eqnarray}
T^{p0}_1& =&- 4m\biggl((k\xi)\Bigl({3\over u_0w_0}+{1\over w_0^2}\Bigr)+{2(q\xi)\over u_0w_0}\biggr),
\nonumber\\
T^{p0}_2& =&- 4 m (k\xi)\biggl({1\over u_0w_0}-{1\over w_0^2}\biggr),
\nonumber\\
T^{p0}_3& =&8 m \Bigl({Q^2\over u_0w_0}-m^2\biggl({1\over u_0}-{1\over w_0}\Bigr)^2\biggr).
\nonumber
\end{eqnarray}
If only leading and next-to-leading terms are kept, only two terms survive in (\ref{leptonicpol0}), and the tensor is reduced to
\begin{eqnarray}
L^{pol}_{\mu\nu}&=&-{2\over u_0w_0} \Bigl(\epsilon_{\mu \nu k q}(u_0+w_0)
+\epsilon_{\mu\nu n q}(t+Q^2)\Bigr).
\end{eqnarray}
or, equivalently, to 
\begin{eqnarray}\label{leptonicpol}
L^{pol}_{\mu\nu}&=&\epsilon_{\mu \nu k q}T^0_5
+\epsilon_{\mu\nu n q}T_6^0,
\end{eqnarray}
where
\begin{eqnarray}
&&  T^0_5=-2\biggl(\frac{1}{w_0}+\frac{1}{u_0}\biggr), \nonumber \\
&&  T^0_6=-2\frac{t+Q^2}{u_0w_0}.
\end{eqnarray}

\section{\label{Loops} One-loop effects}

\begin{figure}[t]
\centering
\scalebox{0.2}{\includegraphics{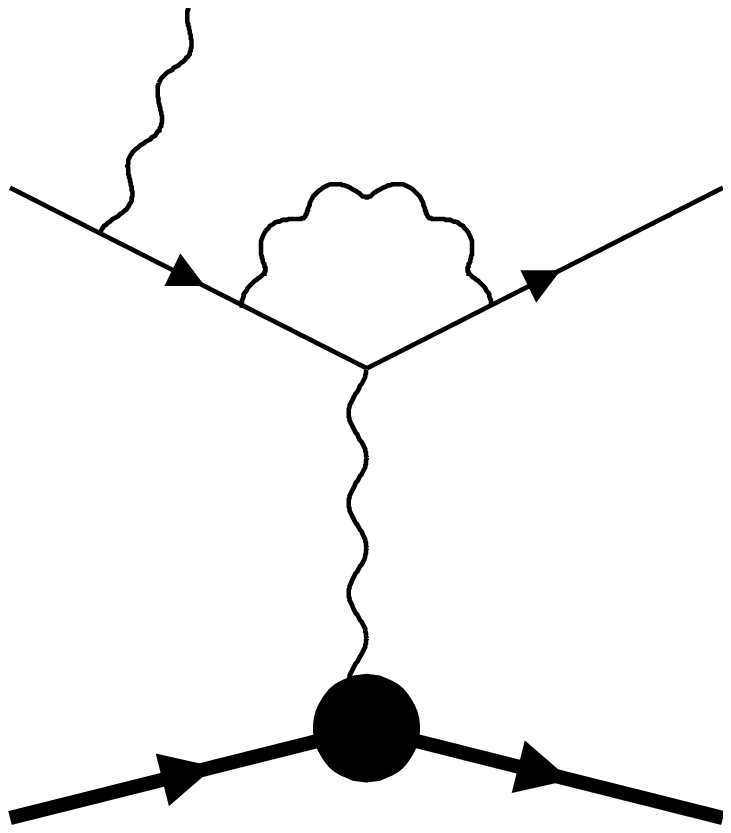}}
\hspace{0.25cm}
\scalebox{0.2}{\includegraphics{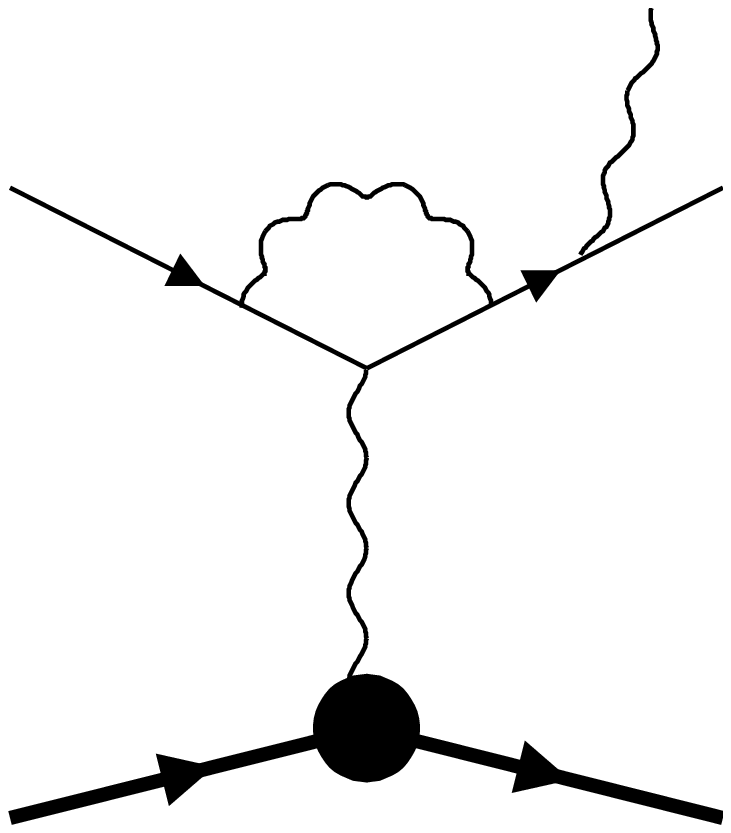}}
\hspace{0.25cm}
\scalebox{0.2}{\includegraphics{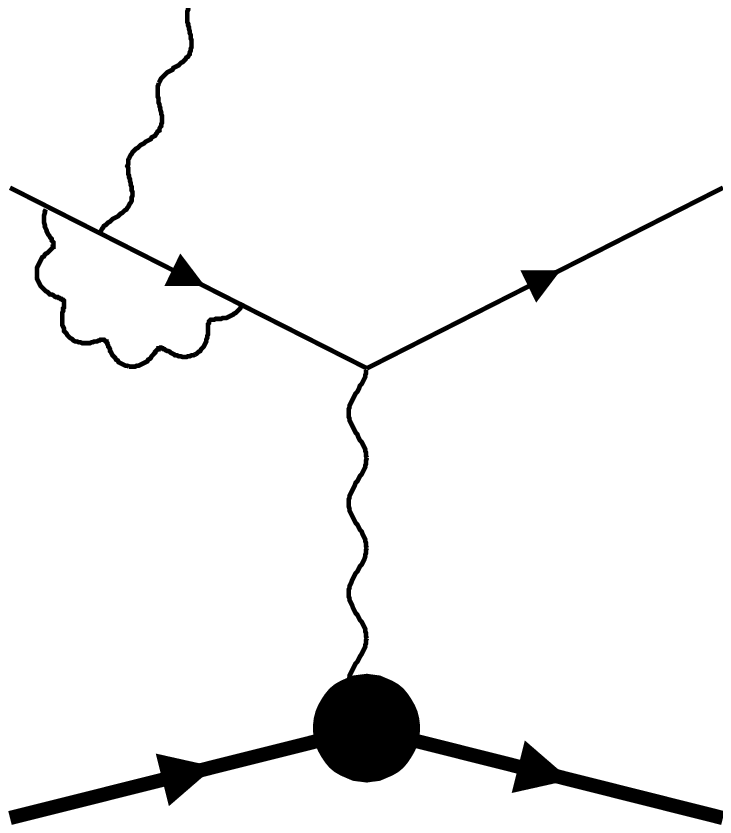}}
\hspace{0.25cm}
\scalebox{0.2}{\includegraphics{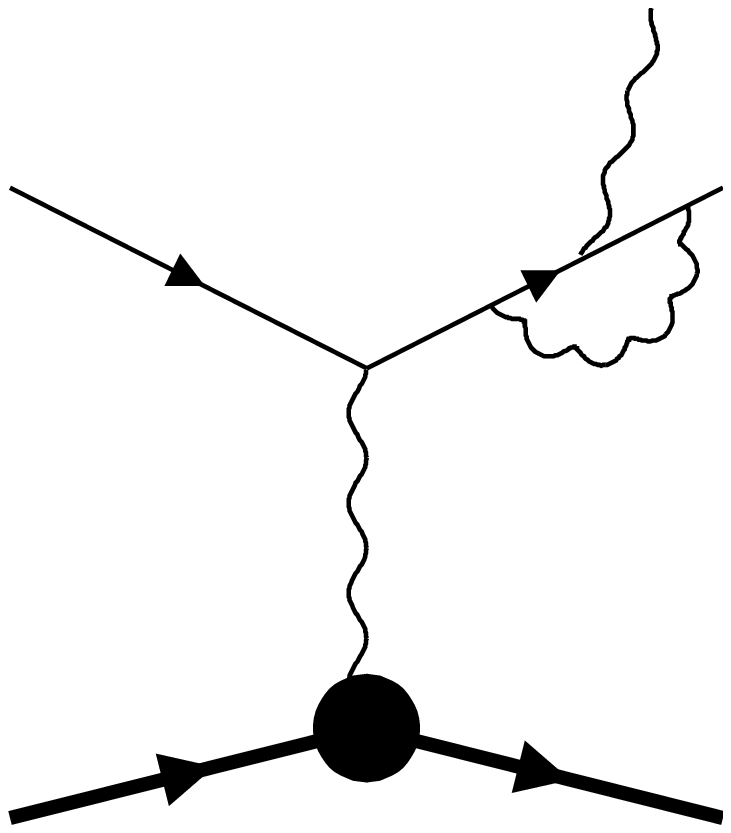}}
\\[-0.1cm]
{\bf \small a) \hspace{1.52cm} b) \hspace{1.52cm} c)\hspace{1.52cm} d)}
\\[0.1cm]
\scalebox{0.2}{\includegraphics{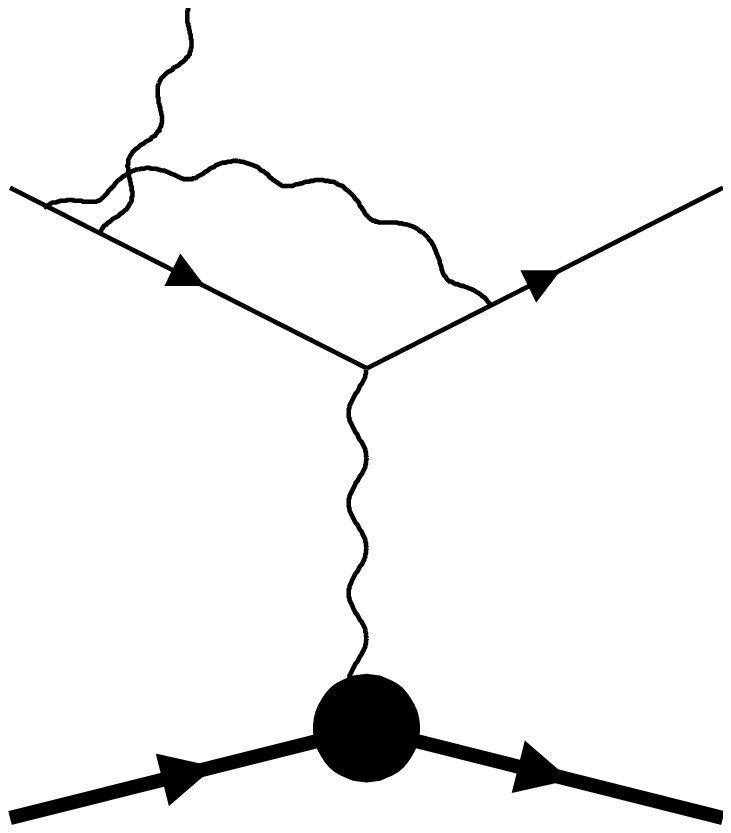}}
\hspace{0.25cm}
\scalebox{0.2}{\includegraphics{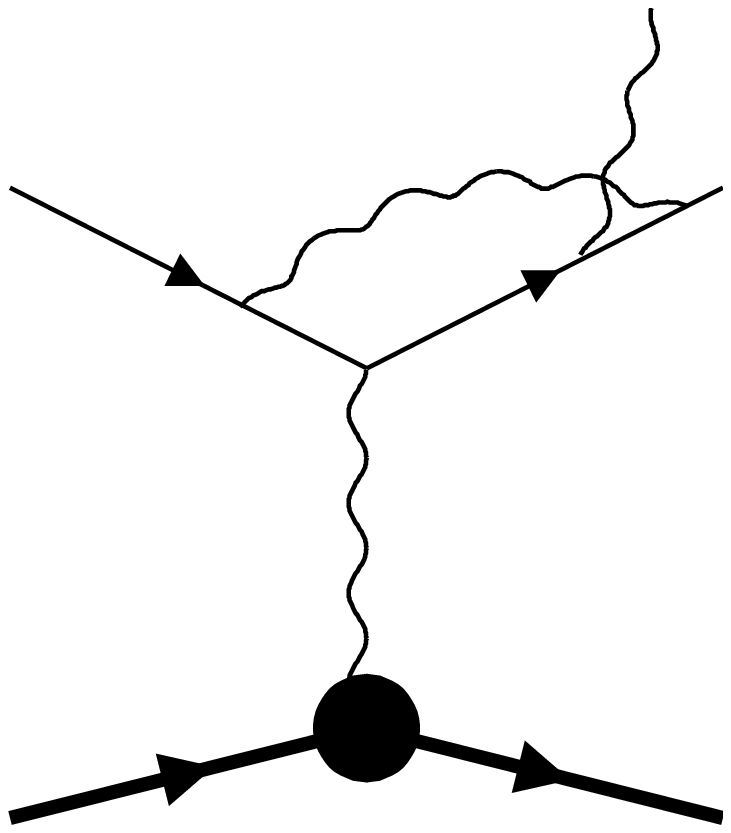}}
\hspace{0.25cm}
\scalebox{0.2}{\includegraphics{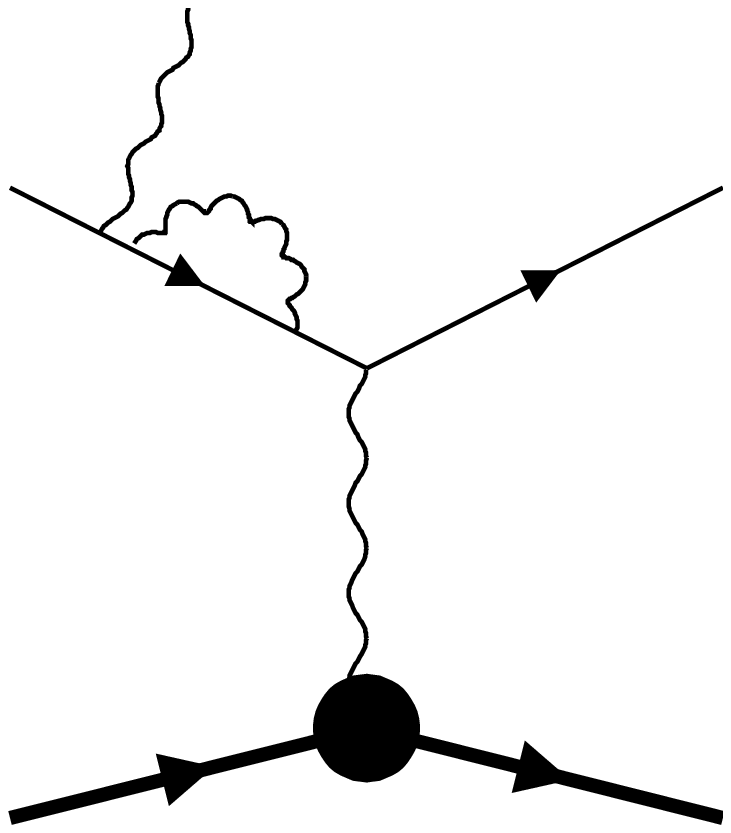}}
\hspace{0.25cm}
\scalebox{0.2}{\includegraphics{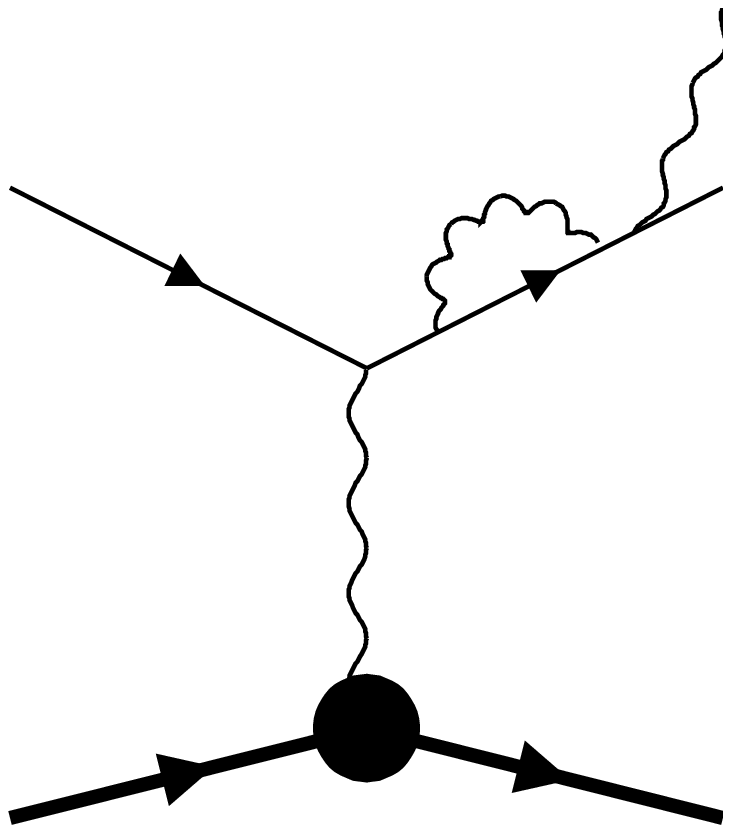}}
\\[-0.1cm]
{\bf \small e) \hspace{1.52cm} f) \hspace{1.52cm} g)\hspace{1.52cm} h)}
\\[0.1cm]
\scalebox{0.18}{\includegraphics{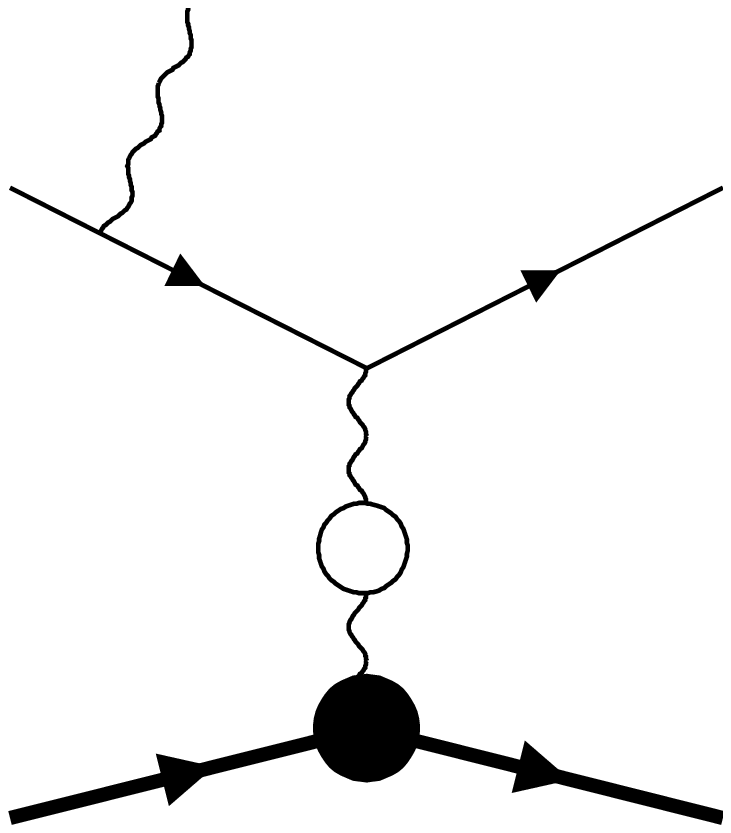}}
\hspace{0.25cm}
\scalebox{0.18}{\includegraphics{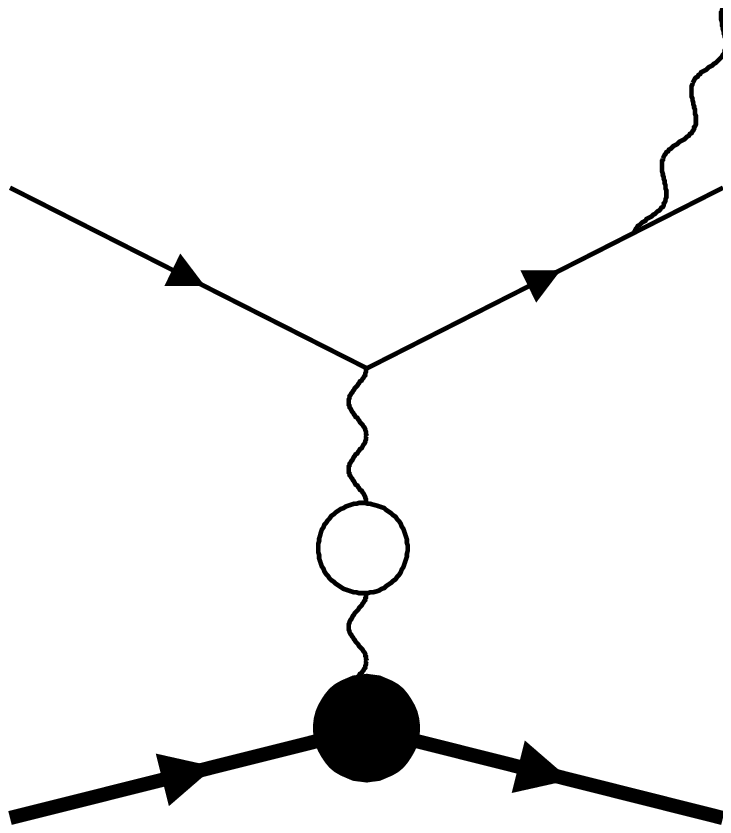}}
\\[-0.1cm]
{\bf \small i) \hspace{1.5cm} j)}
\caption{\label{VVgraphs}Feynman graphs of one-loop effects for the BH cross section}
 \end{figure}

The Feynman diagrams of loops that need to be calculated are presented in Fig. \ref{VVgraphs}. The matrix elements of the eight diagrams with the radiated virtual photon from the lepton line are 
${\cal M}_{loop}=\sum_n (-ie^2/(2\pi)^4) e^3 t^{-1}J^h_\mu J_{n\mu}^{v}$, where
$n$ runs over all graphs presented in Fig. \ref{VVgraphs}{\it a-h}. The
individual $J_{n\mu}^{v}$ are:
\begin{eqnarray}\label{loopJs}
\nonumber
J_{a\mu}^v&=&{\bar u}_2\Gamma^r_\mu(k_1-k,k_2)P(k_1-k){\hat \epsilon} u_1,\\
J_{b\mu}^v&=&{\bar u}_2{\hat \epsilon} P(k_2+k)\Gamma_\mu^r(k_1,k_2+k)u_1,\nonumber\\ 
J_{c\mu}^v&=&{\bar u}_2\gamma_\mu P(k_1-k) \Gamma^r_\alpha (k_1,k_1-k)\epsilon_\alpha u_1,\nonumber\\
J_{d\mu}^v&=&{\bar u}_2\Gamma_\alpha^r(k_2+k,k_2)P(k_2+k)\epsilon_\alpha u_1,\nonumber\\
J_{e\mu}^v&=&{\bar u}_2B_1^{\mu\alpha} \epsilon_\alpha u_1,\nonumber\\
J_{f\mu}^v&=&{\bar u}_2B_2^{\mu\alpha} \epsilon_\alpha u_1,\nonumber\\
J_{g\mu}^v&=&{\bar u}_2\gamma_\mu P(k_1-k) \Sigma^r(k_1-k) P(k_1-k){\hat \epsilon} u_1,\nonumber\\
J_{h\mu}^v&=&{\bar u}_2{\hat \epsilon} P(k_2+k)\Sigma^r(k_2+k)P(k_2+k)\gamma_\mu u_1, 
\end{eqnarray}
where $P$ denotes electron propagator (all factors as $i$, $(2\pi)^4$, and $e$ are combined in front of the expressions for ${\cal M}_{loop}$): 
\begin{equation}
P(a)={{\hat a}+m \over a^2-m^2}
\end{equation}
and
\begin{equation}\label{Boxintegrals}
\begin{array}{l}
\displaystyle
B_1^{\mu\alpha}=\int {d^4l  \over l^2} \gamma_\beta P(k_2-l)\gamma_\mu P(k_1-k-l) \gamma_\alpha P(k_1-l)\gamma_\beta, \nonumber
\\
\displaystyle
B_2^{\mu\alpha}=\int {d^4l \over  l^2} \gamma_\beta P(k_2-l)\gamma_\alpha P(k_2+k-l)\gamma_\mu P(k_1-l) \gamma_\beta.
\end{array}
\end{equation}
$\Gamma^r_\mu(a,a')$ and $\Sigma^r(a)$ are the renormalized vertex functions and the electron self-energy. They are ultraviolet free after combining the respective non-renormalized functions  $\Gamma_\mu(a,a')$ and $\Sigma(a)$ (i.e., obtained formally applying the Feynman rules) and respective counterterms:
\begin{eqnarray}
\Gamma^r_\mu(a,a')&=&\Gamma_\mu(a,a')+\delta_1  \gamma_\mu, 
\nonumber \\
\Sigma^r(a)&=&\Sigma(a)+\delta_m-\delta_2 {\hat a}
\end{eqnarray}
for any vectors $a$ and $a'$. The non-renormalized functions (contributed to matrix elements (\ref{loopJs})) are:
\begin{eqnarray}\label{UVintegrals}
\Gamma_\mu(a,a')&=&\int {d^4l \over l^2}\gamma_\beta P(a'-l)\gamma_\mu P(a-l) \gamma_\beta, \nonumber\\
\Sigma(a)&=&\int {d^4l \over l^2}\gamma_\beta P(a-l) \gamma_\beta.
\end{eqnarray}


The calculation requires the integration over the loop momentum $l$. Because 
of the ultraviolet divergence in (\ref{UVintegrals}) and infrared 
divergence in (\ref{Boxintegrals}) the integrals have to be
regularized first. We use the dimensional regularization in this 
paper allowing for simultaneous dealing with both types of divergences.

The integration over the loop momentum can be performed before or after the calculation of $\gamma$-matrix 
traces. Both  ways have certain limitations 
preventing their straightforward usage in our case. In the latter case, manipulations 
with $\gamma$-matrices has to be performed in the $d$-dimensional space, 
however there are difficulties in the case of polarized particles 
because of ill-defined $\gamma_5$. If integration is performed before 
the calculation of traces, one need to deal with third rank tensors 
(containing $l_\mu l_\nu l_\rho$ in the numerator)
coming from finite part of  (\ref{Boxintegrals}). Therefore we use a combined 
approach in which all integrals in (\ref{UVintegrals}) and the infrared 
divergent part of (\ref{Boxintegrals}), i.e., the part without any 
$l$ in numerators, are calculated in the dimensional regularization. 
The remaining part of (\ref{Boxintegrals}) is infrared free and the 
integrals are calculated for $d=4$ after the calculation of traces. 
In this way all integrals of third rank are reduced to the integrals of lower 
rank. The calculation of the spin-dependent part also requires the 
approximation $\xi=k_1/m$ that is appropriate when calculating with the
leading and next-to-leading accuracy. 

The table of integrals over $l$ is the core in the calculation of 
the loop effects. All integrals can be completely calculated in an 
analytical form.   Two important sources with similar tables are in 
available literature \cite{AB650,Kuraevintegrals}. The tables presented in these 
sources are not complete for our purposes and/or presented in a different 
form. Therefore the table of integrals required for the calculation of  
(\ref{Boxintegrals}) and (\ref{UVintegrals}) was recalculated by us and 
presented in Appendix \ref{loopintegrals}. We followed ref. \cite{Kuraevintegrals} 
in choosing an analytical form for them but used dimensional regularization for all 
divergent terms. 

The integration of (\ref{Boxintegrals}) and (\ref{UVintegrals})  
involves the typical ultraviolet ($P$) and infrared ($P_{IR}$) pole terms that are
defined as in \cite{AB650, AkBaSh1986YP}:
\begin{equation}
P=P_{IR}={1\over d-4}+{\gamma \over 2}+\log\Bigl({m\over
2{\mu}\sqrt{\pi}}\Bigr),
\end{equation}
where $\gamma$ is the Euler constant and $\mu$ is an arbitrary constant of the dimension of  a mass. 
The ultraviolet divergence cancels with respective contributions of the counterterms
\cite{AkBaSh1986YP,Vanderhaeghen2000} that are obtained in the form:
\begin{eqnarray}
\delta_1&=&\delta_2=2(2P_{IR}+P-2), \nonumber\\
\delta_m&=&2(2P_{IR}+4P-4).
\end{eqnarray}

The final result of the calculation  of the  leptonic tensor of the loop effects (graphs in Figure \ref{VVgraphs}a-h) defined as 
\begin{equation}
L_{\mu\nu}^{v}={\alpha\over \pi}{\mathrm Tr} \sum_n \biggl( 
J^{BH}_\mu J_{n\nu}^{v\dagger}
+
J_{n\mu}^{v}J^{BH\dagger}_\nu \biggr)
\end{equation}
can be presented in the form:
\begin{equation}\label{lmunuvv}
L_{\mu\nu}^{v}={\alpha \over \pi}\Bigl( \delta_{IR}^vL_{\mu\nu}^{BH}+\delta_{m}^vL_{\mu\nu}^{BH}+L_{\mu\nu}^{v,unp}+
iL_{\mu\nu}^{v,pol} \Bigr).
\end{equation}
The first term contains the pole infrared terms. It is factorized in front of
the BH leptonic tensor. This term needs to be further manipulated within
$d$-dimensional space. The second term contains the effects of mass
singularities, i.e., dependence on electron mass in the form of $\log m$. It is
also factorized. This term does not contain any pole terms, so it can be treated
in the 4-dimensional space. The explicit form of
the factorized corrections  are
\begin{eqnarray}
\delta_{IR}^v&=&-{1\over 2}\bigl( 4P_{IR}(L_m-1)+1\bigr),
\nonumber \\
\delta_{m}^v&=&-{1\over 2}\bigl( L_m^2-3L_m-\pi^2/3\bigr), 
\label{dirdm}
\end{eqnarray}
where $L_m=\log(Q^2/m^2)$.

Two remaining terms in (\ref{lmunuvv}), i.e., tensors $L_{\mu\nu}^{v,unp}$ and $L_{\mu\nu}^{v,pol}$, represent the next-to-leading
correction. These terms do not contain the electron mass and are different for
unpolarized and spin-dependent parts of the cross section.   
The tensors $L_{\mu\nu}^{v,unp}$ and $L_{\mu\nu}^{v,pol}$ have the same tensor structure as the BH tensors (\ref{leptonicunp}) and (\ref{leptonicpol}): 
\begin{eqnarray}
L^{unp}_{\mu\nu}&=&\tilde{g}_{\mu\nu}T_1^v
+\tilde{k}_\mu \tilde{k}_\nu T_2^v
+\tilde{n}_\mu \tilde{n}_\nu T_3^v
\nonumber\\&&
+(\tilde{k}_\mu \tilde{n}_\nu+\tilde{n}_\mu \tilde{k}_\nu) T_4^v
\end{eqnarray}
and
\begin{eqnarray}\label{leptonicpol56}
L^{pol}_{\mu\nu}&=&
\epsilon_{\mu \nu k q}T^{v}_5
+\epsilon_{\mu\nu n q}T^{v}_6.
\end{eqnarray}
Every $T^v_i$ has the same structure
\begin{eqnarray}\label{ttvv}
T^v_i&=&T_{i0}^v+T_{i1}^vP_{yt}+T_{i2}^vP_w+T_{i3}^vP_u+T_{i4}^vF_i^y
\nonumber\\&&\qquad\qquad
+T_{i5}^vF_i^w+T_{i6}^vF_i^u,
\end{eqnarray}
where functions $F_i^{y,u,w}$ depend on $i$, i.e., $F_i^y=F_{1y}$ for $i$=1,3,6 and $F_i^y=F_{2y}$ for $i$=2,4,5; while $F_i^{w,u}=F_{1w,u}$ for $i$=1 and $F_i^{w,u}=F_{2w,u}$ for $i$=2,3,4,5,6. 
$T_{ij}^v$ in (\ref{ttvv}) are rational functions of $w_0$, $u_0$, and $t$ (or $Q^2=t+u_0-w_0$). The explicit expressions for them are presented in Appendix \ref{Vresults}. Other functions in (\ref{ttvv}) are
\begin{eqnarray}
P_{yt}&=&{\pi^2 \over 3} + 2 \Phi\Bigl(1 - {t \over Q^2} \Bigr) - \log^2{t\over Q^2},
\nonumber\\
P_w&=&\log{t\over Q^2}\log{w_0\over Q^2} - \Phi\Bigl(1 - {w_0 \over t} \Bigr), 
\nonumber\\
P_u&=&\log{t\over Q^2}\log{u_0\over Q^2} - \Phi\Bigl(1 + {u_0 \over t} \Bigr), 
\nonumber\\
F_{1y}&=&{1\over t_y} \log {t\over Q^2},
\nonumber\\
F_{1w}&=&{1\over t_w} \log {t\over w_0},
\nonumber\\
F_{1u}&=&{1\over t_u} \log {u_0\over t},
\nonumber\\
F_{2y}&=&{t F_{1y}-1\over t_y},
\nonumber\\
F_{2w}&=&{w F_{1w}-1\over t_w},
\nonumber\\
F_{2u}&=&{u F_{1u}-1\over t_u},
\end{eqnarray}
where $t_y=t-Q^2=w_0-u_0$, $t_w=t-w_0$, $t_u=t+u_0$, and $\Phi(x)$ is the Spence function defined as
$$
\Phi(x)=-\int\limits_0^x\frac{\log|1-t|}{t}dt.
$$

The remaining one-loop contribution represents the effect of vacuum polarization and is given by graphs in Figure 
\ref{VVgraphs}($i,j$). The contribution is factorized: 
\begin{equation}
L_{\mu\nu}^{vac}=2\Pi (t) L_{\mu\nu}^{BH}={\alpha \over \pi}\delta_{vac}L_{\mu\nu}^{BH},
\end{equation}
where the polarization operator has the standard form given in \cite{AkSh1994}. 

\section{\label{RC}Double bremsstrahlung cross section}

\begin{figure}\centering
\scalebox{0.24}{\includegraphics{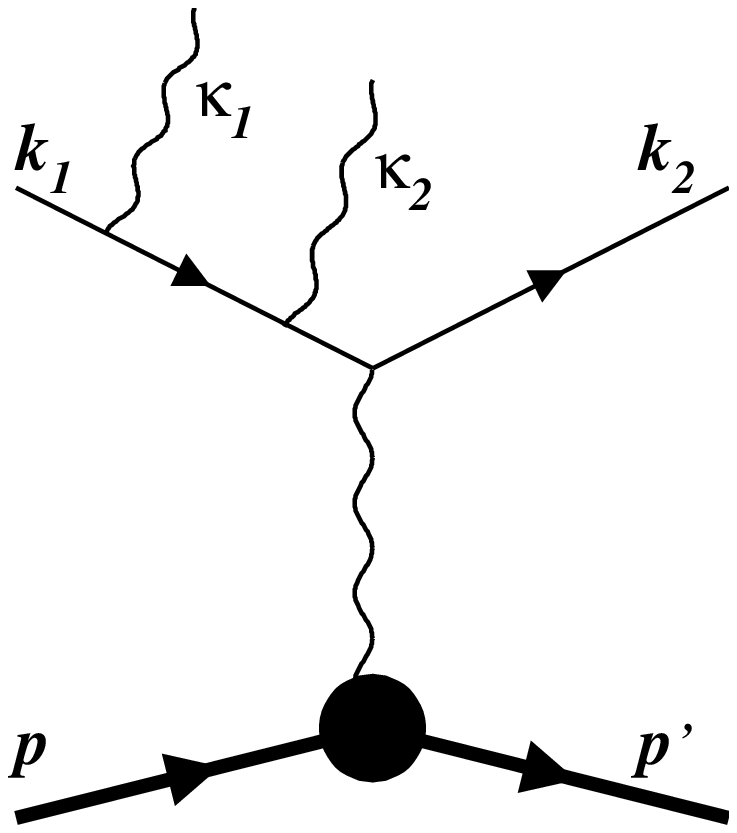}}
\hspace{0.25cm}
\scalebox{0.24}{\includegraphics{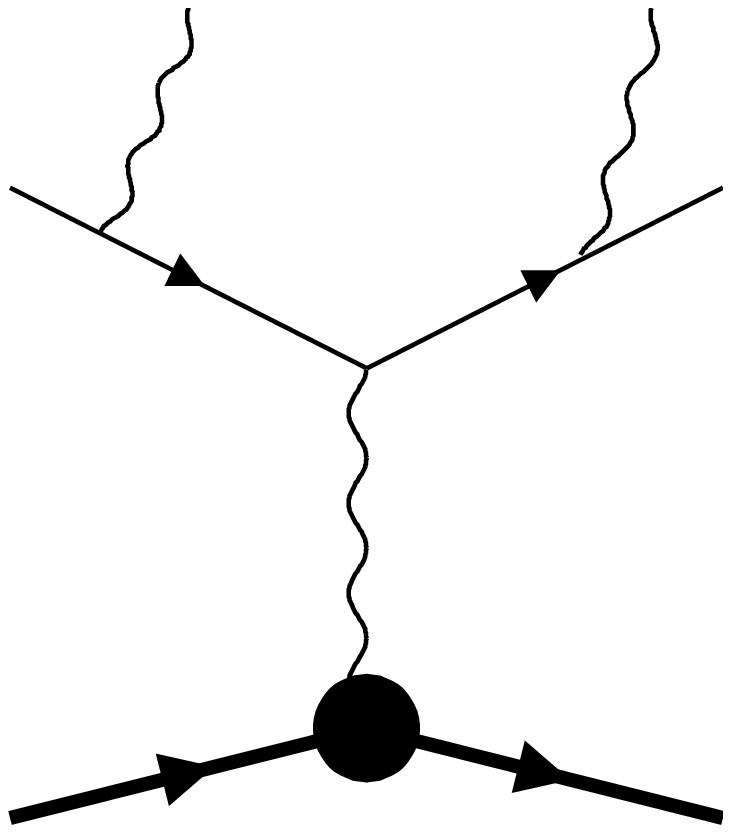}}
\hspace{0.25cm}
\scalebox{0.24}{\includegraphics{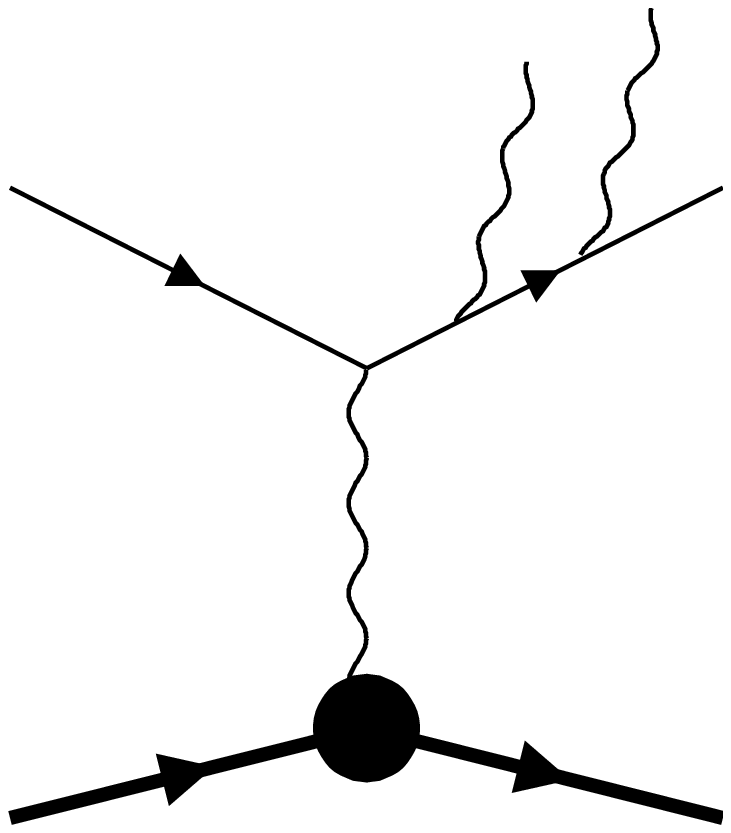}}
\\ 
{\bf a) \hspace{1.8cm} b) \hspace{1.8cm} c)}
\caption{\label{Twoggraphs}Feynman graphs of two real photon emission}
 \end{figure}

The cross section of two photon emission, i.e., the process
\begin{equation}\label{twogammaprocess}
e(k_1,\xi )+p(p,\eta )\longrightarrow e'(k_2)+p'(p')+\gamma(\kappa_1)+\gamma(\kappa_2),
\end{equation}
is
\begin{eqnarray}\label{sigma}
d\sigma&=&\frac{1}{4S} \biggl(\sum_{i=1}^6{\cal M}_i\biggr)^2 d\Gamma,
\end{eqnarray}
where additional factor 2 in the denominator is because of two identical particles (photons) in the final state. 

Phase space is parametrized as \cite{AkBaSh1986YP}:   
\begin{eqnarray}
d\Gamma&=&\frac 1{(2\pi )^8}
\frac{d^3k_2}{2E_2}
\frac{d^3p'}{2p'_0}
\frac{d^3\kappa_1}{2\omega_1}
\frac{d^3\kappa_2}{2\omega_2}
\nonumber 
\\&&\quad\times
\delta^4(k_1+p-k_2-p'-\kappa_1-\kappa_2)
\nonumber 
\\&=&\frac 1{(2\pi)^3}d\Gamma_0dV^2d\Gamma_{2\gamma},
\end{eqnarray}
where  $V^2=\kappa^2=(\kappa_1+\kappa_2)^2$ is the invariant mass of two photons, and $d\Gamma_0$ is given by (\ref{dGamma0}); $\kappa$ denotes four-momentum of the ``heavy'' photon, i.e., the photon with the mass $\sqrt{V^2}$.  Thus, in this parametrization the phase space element for two photons is factorized into that of BH process and an additional phase space $dV^2d\Gamma_{2\gamma}$, where 
\begin{eqnarray}\label{dg2g}
d\Gamma_{2\gamma}&=&{d\kappa_1 \over 2\omega_1}{d\kappa_2 \over 2\omega_2}\delta(\kappa-\kappa_1-\kappa_2)
\nonumber
\\&=&
{1\over 8} d\Omega_R 
={1\over 8} d\cos\theta_Rd\phi_R.
\end{eqnarray}
The angles $\theta_R$ and $\phi_R$ define the orientation of momenta of photons in the system where ${\boldsymbol \kappa}=0$, i.e., in the two-photon central mass system. The
energies of the photons are equal in this system and equal  $V/2$.
Integration over $dV^2d\Gamma_{2\gamma}$ needs to be performed to obtain the RC to the BH cross section (\ref{dsigma0}). 

The maximal value of the invariant mass of two photons $V^2_m$ is defined by kinematics or can be affected by experimental cuts. In the former case the expression for $V^2_m$ is
\begin{equation}\label{vmaxkin}
V^2_m=
{\sqrt{\lambda_t\lambda_Y}-S_xt \over 2M^2}-Q^2-t,	
\end{equation}
where $\lambda_t=t^2+4tM^2$.


Six matrix elements of the process with emission of an additional photon correspondent 
to graphs in Figure \ref{Twoggraphs} are denoted ${\cal M}_{1-6}=e^4t^{-1}J_\mu^h J_{1-6,\mu}$. The quantities $J_{1-6,\mu}$, proportional to the leptonic currents, are:
\begin{eqnarray}
J_{1\mu}&=& 
{\bar u}_2 
\gamma_\mu 
\frac{{\hat k}_1-{\hat \kappa}+m}{-2\kappa k_1+V^2}
{\hat \epsilon}_2
\frac{{\hat k}_1-{\hat \kappa}_1+m}{-2k_1\kappa_1}
{\hat \epsilon}_1
u_1,
\nonumber \\
J_{2\mu}&=& 
{\bar u}_2 
\gamma_\mu 
\frac{{\hat k}_1-{\hat \kappa}+m}{-2\kappa k_1+V^2}
{\hat \epsilon}_1
\frac{{\hat k}_1-{\hat \kappa}_2+m}
{-2k_1\kappa_2}{\hat \epsilon}_2
u_1,
\nonumber \\
J_{3\mu}&=& 
{\bar u}_2 
{\hat \epsilon}_2
\frac{{\hat k}_2+{\hat \kappa_2}+m}{2k_2\kappa_2}
{\hat \epsilon}_1
\frac{{\hat k}_2+{\hat \kappa }+m}{2\kappa k_2+V^2}
\gamma_\mu 
u_1,
\nonumber \\
J_{4\mu}&=& 
{\bar u}_2 
{\hat \epsilon}_1
\frac{{\hat k}_2+{\hat \kappa_1}+m}{2k_2\kappa_1}
{\hat \epsilon}_2
\frac{{\hat k}_2+{\hat \kappa }+m}{2\kappa k_2+V^2}
\gamma_\mu 
u_1,
\nonumber \\
J_{5\mu}&=& 
{\bar u}_2 
{\hat \epsilon}_1
\frac{{\hat k}_2+{\hat \kappa_1}+m}{2k_2\kappa_1}
\gamma_\mu 
\frac{{\hat k}_1-{\hat \kappa}_2+m}{-2k_1\kappa_2}
{\hat \epsilon}_2
u_1,
\nonumber \\
J_{6\mu}&=& 
{\bar u}_2 
{\hat \epsilon}_2
\frac{{\hat k}_2+{\hat \kappa_2}+m}{2k_2\kappa_2}
\gamma_\mu 
\frac{{\hat k}_1-{\hat \kappa}_1+m}{-2k_1\kappa_1}
{\hat \epsilon}_1
u_1.
\end{eqnarray}
The total leptonic current is
\begin{equation}
J_{\mu}^{2\gamma}=\sum_{i=1}^6  J_{i\mu}.
\end{equation}
The matrix element squared can be represented in terms of leptonic and hadronic tensors. The former is defined as 
\begin{eqnarray}
L_{\mu\nu}(\kappa_1,\kappa_2)&=&{\rm Tr}\sum J_{\mu}^{2\gamma}
J_{\nu}^{2\gamma\dagger}
\end{eqnarray}
and the latter is given by (\ref{wmunu}-\ref{wmunupol}).
The sum means averaging over polarization states of initial lepton and summing over polarization states of photons and final leptons.
Since the dependence on $\kappa_1$ and $\kappa_2$ are only contained in $L_{\mu\nu}(\kappa_1,\kappa_2)$, the cross section (\ref{sigma})  can be presented as
\begin{equation}\label{sigmab}
{d\sigma \over d\Gamma_0}={8\pi \alpha^4 \over St^2} \int dV^2 
W_{\mu\nu}
\int d\Gamma_{2\gamma} L_{\mu\nu}(\kappa_1,\kappa_2) .
\end{equation}
The integration over $d\Gamma_{2\gamma}$ involving scalar, vector, and tensor integrals can be performed analytically. 

The cross section (\ref{sigmab}) contains the infrared divergence, and therefore, has to be calculated in the $d$-dimensional space. The extraction of the infrared divergence from $ L_{\mu\nu}(\kappa_1,\kappa_2)$ is based on observation that it is contained only in terms 
\begin{equation}\label{FF12}
(F_1^{IR}+F_2^{IR})L_{\mu\nu}^{BH}(\kappa),
\end{equation}
where 
\begin{equation}\label{F12IR}
F_{1,2}^{IR}=\biggl( {k_2\over u_{1,2}}-{k_1 \over w_{1,2}}\biggr)^2={Q^2+2m^2\over u_{1,2}w_{1,2}}-{m^2\over u_{1,2}^2}-{m^2\over w_{1,2}^2}.
\end{equation}
Here $w_{1,2}=2k_1\kappa_{1,2}$ and $u_{1,2}=2k_2\kappa_{1,2}$. 
Integrals $\int d\Gamma_{2\gamma} F_{1,2}^{IR}$ are proportional to $1/V^2$, therefore the external integral over $V^2$ has the infrared divergence for $V^2=0$. $L_{\mu\nu}^{BH}(\kappa)$ in (\ref{FF12}) is the leptonic tensor describing emission of the ``heavy'' photon with the momentum $\kappa$ and the mass $\sqrt{V^2}$: 
\begin{eqnarray}
L_{\mu\nu}^{BH}(\kappa)&=&\tilde{g}_{\mu\nu}T_1^\kappa
+\tilde{\kappa}_\mu \tilde{\kappa}_\nu T_2^\kappa
+\tilde{n}_\mu \tilde{n}_\nu T_3^\kappa
\nonumber\\&&
+(\tilde{\kappa}_\mu \tilde{n}_\nu+\tilde{n}_\mu \tilde{\kappa}_\nu) T_4^\kappa
\nonumber\\&&
+i\Bigl(\epsilon_{\mu \nu k q}T_1^{p\kappa}
+\epsilon_{\mu\nu n q}T_2^{p\kappa}\Bigr),
\end{eqnarray}
where
\begin{eqnarray}
T_1^\kappa&=&-2\biggl( {V_{\mathrm I}\over z_1^2}+{V_{\mathrm I}\over z_2^2} +{2tQ^2\over z_1 z_2}\biggr),
\nonumber\\
T_2^\kappa&=&-2\biggl( {2w-V^2\over z_1^2}-{2u+V^2\over z_2^2} -{2Q^2\over z_1 z_2} \biggr),
\nonumber\\
T_3^\kappa&=&-2\biggl( {V^2\over z_1^2}+{V^2\over z_2^2} -{2(t+V^2)\over z_1 z_2}\biggr),
\nonumber\\
T_4^\kappa&=&-2\biggl( -{w\over z_1^2}-{u\over z_2^2} +{w+u\over z_1 z_2} \biggr),
\nonumber\\
T_1^{p\kappa}&=&-2\biggl({w\over z_1^2}+{u\over z_2^2}  \biggr),
\nonumber\\
T_2^{p\kappa}&=&-2\biggl( -{V^2\over z_1^2}-{V^2\over z_2^2} -{(Q^2+t-V^2)\over z_1 z_2}\biggr).
\label{T123456}
\end{eqnarray}
Here $z_1=w-V^2$, $z_2=u+V^2$, $V_{\mathrm I}=uw-Q^2V^2$, and
\begin{eqnarray}\label{uwexplicit}
\{w,u\}&=&\pm\frac{1}{2}(t+V^2-Q^2)+{S_p\over 2\lambda_Y}\bigl(S_x(Q^2+t+V^2)
\nonumber\\&&\quad
-2tQ^2\bigr)
+{\sqrt{\lambda^v_{uw}} \over \lambda_Y} \cos\phi,
\end{eqnarray}
where 
\begin{eqnarray}\label{lambdav}
\lambda^v_{uw}&=&4(SXQ^2-M^2Q^4-m^2\lambda_Y)
\nonumber\\&&
\times\bigl((S_{xt}(S_x-Q^2)-S_xV^2)t
\nonumber\\&&
-M^2((Q^2-t-V^2)^2+4Q^2V^2)\bigr).
\end{eqnarray}
The expressions (\ref{uwexplicit}) and (\ref{lambdav}) for 
$w=2\kappa k_1$ and $u=2\kappa k_2$ generalize (\ref{uwexplicit00}) and (\ref{lambdav00}) for the case of the ``heavy'' photon.
Note that $u=w+Q^2-t-V^2$.

The representation of
$L_{\mu\nu}(\kappa_1,\kappa_2)$ in terms of three summands is used to extract the infrared divergence
\begin{eqnarray}\label{threesummands}
L_{\mu\nu}(\kappa_1,\kappa_2)&=&L_{\mu\nu}^F(\kappa)+4(F_1^{IR}+F_2^{IR})
\bigl( L_{\mu\nu}^{BH}(\kappa)-L_{\mu\nu}^{BH} \bigr)
\nonumber\\&& 
+4(F_1^{IR}+F_2^{IR})L_{\mu\nu}^{BH}.
\end{eqnarray}
The first summand is the tensor $L_{\mu\nu}(\kappa_1,\kappa_2)$ without infrared terms (\ref{FF12}):
\begin{equation}
L_{\mu\nu}^F(\kappa)=L_{\mu\nu}(\kappa_1,\kappa_2)
-4(F_1^{IR}+F_2^{IR})L_{\mu\nu}^{BH}(\kappa).
\end{equation}
It is infrared free, and therefore, $d=4$ can be used for its calculation. The second summand is also infrared free because $L_{\mu\nu}^{BH}(\kappa)\rightarrow L_{\mu\nu}^{BH}$ and $\kappa\rightarrow k$ for $V^2\rightarrow 0$. $d=4$ is used for this summand too. 

The infrared divergence is contained in third summand that contributes to the cross section (\ref{sigmab}) in the form \cite{AkBaSh1986YP,ABMO1985}:
\begin{equation}
{d\sigma_{IR} \over d\Gamma_0}={\alpha \over \pi} \delta_R^{IR} {d\sigma_{0} \over d\Gamma_0}
\end{equation}
and
\begin{equation}
  \delta_R^{IR}={1\over 4\pi} \int\limits_0^{V_m^2} dV^2 \int d\Gamma_{2\gamma} 4(F^{IR}_{1}+F^{IR}_{2}).
\end{equation}
The integration over the 3-momentum of one of photons and then to over $V^2$ is performed using the $\delta$-function in (\ref{dg2g}). The integration region over the momentum of remaining photon (denoted by $\kappa_{cm}$) in the two-photon center-mass system  can be splited into two parts by an infinitesimal parameter ${\bar \kappa}$ resulting in $\delta_R^{IR}=\delta_1+\delta_2$ with
  \begin{eqnarray}
  \delta_1&=&{1\over 4\pi} \int\limits_0^{V_m^2} dV^2 \int d\Gamma_{2\gamma} 4(F^{IR}_{1}+F^{IR}_{2})
 \theta({\bar \kappa}-\kappa_{cm}),
\nonumber \\
  \delta_2&=&{1\over 4\pi} \int\limits_0^{V_m^2} dV^2 \int d\Gamma_{2\gamma} 4(F^{IR}_{1}+F^{IR}_{2})
 \theta(\kappa_{cm}-\bar \kappa).
  \end{eqnarray}
  The second term does not contain infrared divergence and is calculated straightforwardly 
  \begin{equation}
  \delta_2=2\log\Bigl({V^2_m\over 4{\bar \kappa}^2}\Bigr)(L_m-1),
  \label{del2}
  \end{equation}
where $L_m$ is defined after eq. (\ref{dirdm}).

  The calculation of the first term is performed in the dimensional regularization. 
  The phase space of the remaining photon (after integration using the $\delta$-function) is rewritten in the $d$-dimensional space as in \cite{BardinShumeiko1977} (see eq. 47) resulting in the expression for $\delta_1$ in the form: 
  \begin{eqnarray}
  \delta_1&=&
 {(2\sqrt{\pi}\mu)^{4-d} \over \Gamma(d/2-1)}
  \int\limits_{0}^1 d\alpha\
  \int\limits_{0}^{\bar\kappa} {d\kappa_{cm} \over \kappa_{cm}^{5-d}}
  \int\limits_{-1}^1 d\zeta 
  (1-\zeta^2)^{\frac{d}{2}-2}
  \nonumber
  \\&&
  \times\biggl({Q^2+2m^2\over (E_\alpha-p_\alpha \zeta)^2}-{m^2\over  (E_1-p_1 \zeta)^2}
  \nonumber
  \\&&
  -{m^2\over  (E_2-p_2 \zeta)^2}\biggr),
 \end{eqnarray}
  where $\mu$ is an arbitrary parameter of the dimension of a mass. The energies of the initial and final electrons $E_1$ and $E_2$ are taken in the two-photons center-mass system: 
\begin{equation}
  E_1={w\over 4k_{cm}}, \quad
  E_2={u\over 4k_{cm}}, \quad
  E_\alpha={w\alpha +u(1-\alpha ) \over 4k_{cm}}
 \end{equation}
  and $p_\alpha^2=E_\alpha^2-m_\alpha^2$, $m_\alpha^2=m^2+\alpha(1-\alpha)Q^2$. 
  The first step in the calculation is the integration over $\zeta$. The result of this integration involves the hyperheometric function, however allows for expansion over $k_{cm}$. The forthcoming integration over $k_{cm}$, extraction of infrared divergent terms, integration over $\alpha$, and expansion over 
  $m$ keeping only leading and next-to-leading terms result in
\begin{eqnarray}
  \delta_R^{IR}&=&\biggl(2 P_{IR} + \log\Bigl({V^4_m \over u_0w_0}\Bigr)\biggr) (L_m - 1) + {1 \over2 }L^2_m 
  \nonumber\\&&\qquad
  - {\pi^2\over 6}  - {1\over 2} \log^2{u_0\over w_0}.
  \end{eqnarray}

%

  Integrals over $d\Gamma_{2\gamma}$ for infrared free terms (i.e., first and second summands in (\ref{threesummands})) can be performed in the 4-dimensional space. All integrals are calculated in Appendix \ref{Rintegrals}. For the second summand the integration results in
\begin{eqnarray}  
\int d\Gamma_{2\gamma} 4(F^{IR}_{1}+F^{IR}_2)= {8\pi \over V^2} (L_m-1)+O\biggl (\frac{m^2}{Q^2}\biggr).
\end{eqnarray}

The integration of the first summand in (\ref{threesummands}) is cumbersome resulting in
\begin{eqnarray}\label{lmunu2gamma}
L_{\mu\nu}^{F}(\kappa)&=&\tilde{g}_{\mu\nu}T_1^F
+\tilde{\kappa}_\mu \tilde{\kappa}_\nu T_2^F
+\tilde{n}_\mu \tilde{n}_\nu T_3^F
\nonumber\\&&
+(\tilde{\kappa}_\mu \tilde{n}_\nu+\tilde{n}_\mu \tilde{\kappa}_\nu) T_4^F
\nonumber\\&&
+i\Bigl(\epsilon_{\mu \nu k q}T_5^F
+\epsilon_{\mu\nu n q}T_2^6\Bigr).
\end{eqnarray}
All $T_i^F$ have the same structure
\begin{eqnarray}\label{TiF}
\frac{1}{4\pi}T_i^F&=&
{T^F_{i1}\over w}\log{w^2\over m^2V^2}
+{T^F_{i2}\over u}\log{u^2\over m^2V^2}
+T^F_{i3}\log{V_{\mathrm I}\over m^2V^2}
\nonumber\\&&
+T^F_{i4}\log{Q^2\over m^2}
+T^F_{i5}.
\end{eqnarray}
Quantities $T^F_{ij}$ are rational functions of $Q^2$, $u$, $w$, and $V^2$. They are independent of the electron mass. The explicit expressions for the Quantities $T^F_{ij}$ are presented in Appendix \ref{Rappendix}. There are four remarks worth discussion. 

First, as one can see many integrals over $\Gamma_{2\gamma}$ contain $m^2$ in denominators and therefore are infinite in the massless approximation (see Appendix \ref{Rintegrals}). However all these terms completely cancel when combined into functions $T^F_{ij}$ and these functions are finite in the massless approximation. 

Second, the lepton polarization vector was used in the form 
\begin{equation}\label{xixi} 
\xi={k_1\over m}+m\bar \xi,
\end{equation}
where $\bar \xi$ does not result in any lepton mass dependence. This corresponds to the most practical case of longitudinally polarized lepton or more general case when lepton polarization vector is represented in the form of (\ref{xixi}). 
After integration over  $\Gamma_{2\gamma}$ the vector $\bar\xi$ appeared in $T_{ij}^F$ only in the form $k_1{\bar\xi}=-1$.
This is a reason why $\bar \xi$ does not explicitly contribute to (\ref{TiF}) and equations for $T^F_{ij}$ given in Appendix \ref{Rappendix}. 

Third, using known properties of traces of $\gamma$-matrices one can show that the spin-independent part and spin-dependent part containing leading component of $\xi\approx k_1/m$ of $L_{\mu\nu}^{F}(\kappa)$ are symmetric in respect of substitution $k_1 \leftrightarrow k_2$ and $\kappa \rightarrow -\kappa$. 
Therefore functions $T^F_{ij}$ possess certain symmetry (i.e., symmetric or antisymmetric dependent on the symmetry of respective tensors), for example,  $T^F_{i1}\leftrightarrow T^F_{i2}$ and $T^F_{i3,4}\leftrightarrow T^F_{i3,4}$
for $i$=1, 2, 3, and 6, while $T^F_{i1}\leftrightarrow -T^F_{i2}$ and $T^F_{i3,4}\leftrightarrow -T^F_{i3,4}$ for $i$=4 and 5. Because of the contribution of $\bar \xi$, functions $T^F_{i5}$ have certain symmetry only for unpolarized part of the leptonic tensor. 

Fourth, the results obtained earlier in the leading log approximation \cite{AkushevichIlyichev2012}, when only terms containing $\log m^2$ are held, are directly obtained from $T^F_i$ presented in  
(\ref{TiF}) and Appendix \ref{Rappendix}. The original leptonic tensor $L_{\mu\nu}(\kappa_1,\kappa_2)$ integrated over $\Gamma_{2\gamma}$ (denoted by $\int d\Gamma_{2\gamma}L_{\mu\nu}(\kappa_1,\kappa_2)\equiv L_{\mu\nu}^\kappa$) can be represented as  
\begin{equation}\label{LmunuLL}
L_{\mu\nu}^\kappa={4L_m \over w}{1+z_1^2\over z_1 (1-z_1)} L^s_{\mu\nu} 
+{4L_m \over u}{1+z_2^2\over 1-z_2} L^p_{\mu\nu},
\end{equation}
where $L^s_{\mu\nu}$ and $L^p_{\mu\nu}$ are leptonic tensors for BH taken in a shifted kinematical points. If to explicitly use arguments of BH leptonic tensor,
\begin{equation}
L_{\mu\nu}^{BH}\equiv L_{\mu\nu}^{BH}(k_1,k_2,k,w,u,Q^2)
\end{equation}
then 
\begin{eqnarray}\label{LLz}
L^s_{\mu\nu}&=&L_{\mu\nu}^{BH}(z_1k_1,k_2,
\kappa-(1-z_1)k_1,
\nonumber\\&&\qquad\qquad
z_1w,u-(1-z_1)Q^2,z_1Q^2),
\nonumber\\
L^p_{\mu\nu}&=&L_{\mu\nu}^{BH}(k_1,{k_2\over z_2},
\kappa-({1\over z_2}-1)k_2,
\nonumber\\&&\qquad\qquad
w-({1\over z_2}-1)Q^2,{u\over z_2},{Q^2\over z_2}).
\end{eqnarray}

\section{\label{ocs123}Cross section}

The hadronic tensor (\ref{wmunu}-\ref{wmunupol}) can be rewritten in the form 
\begin{equation}
W_{\mu\nu}=\sum\limits_{j=1}^4 w^j_{\mu\nu} {\cal F}_j,
\end{equation}
where the structure functions are defined as ($\tau=t/4M^2$) 
\begin{eqnarray}
&&{\cal F}_1=(F_1+F_2)^2,
\quad \quad \quad \quad \quad
 {\cal F}_2=F_1^2+\tau F_2^2,
\nonumber\\
&&{\cal F}_3=(F_1+F_2)(F_1-\tau F_2),
\;\;
{\cal F}_4=(F_1+F_2)F_2
\end{eqnarray}
and 
\begin{eqnarray}\label{wsmall}
&&w_{\mu\nu}^1=-t{\tilde g}_{\mu\nu},
\quad 
\quad
\quad
\;\;
w_{\mu\nu}^2=4{\tilde p}_{\mu}{\tilde p}_{\nu},
\nonumber\\
&&w_{\mu\nu}^3=2iM\epsilon_{\mu\nu \eta q},
\quad \quad
w_{\mu\nu}^4=i{\eta q \over M}\epsilon_{\mu\nu p q}.
\end{eqnarray}

Contractions of the leptonic tensor describing $2\gamma$ emission 
(\ref{lmunu2gamma}) with the tensors contributing to hadronic tensor (\ref{wmunu}) requires the calculation of contraction of the base tensors (i.e., ${\tilde g}_{\mu\nu}$, ${\tilde\kappa}_\mu{\tilde\kappa}_\nu$, etc) with (\ref{wsmall}):
\begin{eqnarray}\label{Cij}
C_{11}&=&{\tilde g}_{\mu\nu}w^1_{\mu\nu}=-3t,
\nonumber\\
C_{21}&=&{\tilde\kappa}_\mu{\tilde\kappa}_\nu w^1_{\mu\nu}
=-{(u-w)^2+4Q^2V^2\over 4},
\nonumber\\
C_{31}&=&{\tilde n}_\mu{\tilde n}_\nu w^1_{\mu\nu}
=-{(u+w)^2+4Q^2t\over 4},
\nonumber\\
C_{41}&=&({\tilde n}_\mu{\tilde\kappa}_\nu+{\tilde n}_\nu{\tilde\kappa}_\mu) w^1_{\mu\nu}
=-{(Q^2+t+V^2)(u+w)\over 2},
\nonumber\\
C_{12}&=&{\tilde g}_{\mu\nu} w^2_{\mu\nu}={t+4M^2},
\nonumber\\
C_{22}&=&{\tilde\kappa}_\mu{\tilde\kappa}_\nu w^2_{\mu\nu}={(2S_x-Q^2-t-V^2)^2\over 4},
\nonumber\\
C_{32}&=&{\tilde n}_\mu{\tilde n}_\nu w^2_{\mu\nu}
={(2S_p-u-w)^2\over 4},
\nonumber\\
C_{42}&=&({\tilde n}_\mu{\tilde\kappa}_\nu+{\tilde n}_\nu{\tilde\kappa}_\mu) w^2_{\mu\nu}
\nonumber\\&=&{(2S_x-Q^2-t-V^2)(2S_p-u-w)\over 2},
\nonumber\\
C_{53}&=&i\epsilon_{\mu\nu k q} w^3_{\mu\nu}
=-2M\bigl(2t(\eta k)-(Q^2-t+V^2)(\eta q)\bigr),
\nonumber\\
C_{63}&=&i\epsilon_{\mu\nu n q} w^3_{\mu\nu}
=-2M\bigl(2t(\eta n)-(u+w)(\eta q)\bigr),
\nonumber\\
C_{54}&=&i\epsilon_{\mu\nu k q} w^4_{\mu\nu}
=-(\eta q){(2S_x-Q^2-t-V^2)t \over 2M}, 
\nonumber\\
C_{64}&=&i\epsilon_{\mu\nu n q} w^4_{\mu\nu}
=-(\eta q){(2S_p-u-w)t \over 2M}.
\end{eqnarray}
Scalar products of polarization vector $\eta=(0,\eta_x,\eta_y,\eta_z)$ with four-momenta are obtained as generalizations of Eqs. (15) of ref. \cite{AkushevichIlyichev2012} for the case of the heavy photon.  
\begin{eqnarray}
\label{etaxyz}
\eta k_1&=&-\sqrt{\lambda_{SX} \over \lambda_Y}\;\eta_x-{SS_x+2M^2Q^2 \over 2M\sqrt{\lambda_Y}}\;\eta_z, 
\nonumber \\
\eta k_2&=&-\sqrt{\lambda_{SX} \over \lambda_Y}\;\eta_x-{XS_x-2M^2Q^2 \over 2M\sqrt{\lambda_Y}}\;\eta_z, 
\nonumber \\
\eta p'&=&-{\sqrt{{\bar \lambda}_{SX}\over \lambda_Y}} (\eta_x\cos\phi+\eta_y\sin\phi)
\nonumber \\&&\quad
-{tS_x+2M^2(Q^2+t+V^2) \over 2M\sqrt{\lambda_Y}}\;\eta_z
\end{eqnarray}
and $\eta k=\eta k_1-\eta k_2-\eta p'$. Here $\lambda_{SX}$ and ${\bar\lambda}_{SX}$ are factors in $\lambda^v_{uw}$ in (\ref{lambdav}) taken in ultrarelativistic approximation:
\begin{eqnarray}
\lambda_{SX}&=&SXQ^2-M^2Q^4,
\nonumber\\
{\bar \lambda}_{SX}&=&
\bigl((S_{xt}(S_x-Q^2)-S_xV^2)t
\nonumber\\&&
-M^2((Q^2-t-V^2)^2+4Q^2V^2)\bigr).
\end{eqnarray}

Contractions of the tensors describing BH and the loop-effect leptonic tensors are denoted by $C_{ij}^0$ and obtained from (\ref{Cij}) using substitutions $w\rightarrow w_0$,
$u\rightarrow u_0$, and $V^2\rightarrow 0$. Thus, the BH cross section is obtained in the form
\begin{eqnarray}\label{sigmaBH000}
{d\sigma_0 \over d\Gamma_0}&=&{32 \pi^3\alpha^3 \over St^2}
\Bigl(
\sum\limits_{i=1}^4 \sum\limits_{j=1}^2
C_{ij}^0 T^0_i {\cal F}_j
\nonumber
\\&&
+P_LP_N
\sum\limits_{i=5}^6 \sum\limits_{j=3}^4
C_{ij}^0 T^0_i {\cal F}_j
\Bigr)
\end{eqnarray}
or briefly
\begin{eqnarray} \label{sigmaBH001}
{d\sigma_0 \over d\Gamma_0}&=&{32 \pi^3\alpha^3 \over St^2}
\sum\limits_{i,j}
C_{ij}^0 T^0_i {\cal F}_j.
\end{eqnarray}
Here $\sum_{i,j}$ denotes the sum over all $i,j$ with non-zero (\ref{Cij}) as in  
(\ref{sigmaBH000}), i.e., 
\begin{equation}
\sum\limits_{i,j}\rightarrow \sum\limits_{i=1}^4 \sum\limits_{j=1}^2
+P_LP_N \sum\limits_{i=5}^6 \sum\limits_{j=3}^4.
\end{equation}

The BH cross section coincides with the cross section for unpolarized and polarized targets given by (7-10) of ref. 
\cite{AkushevichIlyichev2012} as well as with results given in \cite{BMK2002}.
The contribution of loops contains both factorized terms (containing infrared divergent and mass singularity terms) and non-factorized  
\begin{equation}
{d\sigma_v \over d\Gamma_0}=
{\alpha \over \pi}(\delta^v_{IR}+\delta^v_m+\delta_{vac}){d\sigma_0 \over d\Gamma_0}
+{d\sigma_v^F \over d\Gamma_0},
\end{equation}
where
\begin{equation}
{d\sigma_v^F \over d\Gamma_0}=
{32 \pi^3\alpha^3 \over St^2}{\alpha\over \pi}
\sum\limits_{i,j}
C_{ij}^0 T^v_i {\cal F}_j.
\end{equation}

The contribution of two-gamma bremsstrahlung is
\begin{equation}
{d\sigma_R \over d\Gamma_0}=
{\alpha \over \pi}\delta_R^{IR}{d\sigma_0 \over d\Gamma_0}
+{d\sigma_d \over d\Gamma_0}
+{d\sigma_R^F \over d\Gamma_0},
\end{equation}
where
\begin{eqnarray}
{d\sigma_d \over d\Gamma_0}&=&
{64 \pi^2\alpha^4 \over St^2}(L_m-1)
\sum\limits_{i,j} {\cal F}_j
\int\limits_0^{v^2_m} {dV^2\over V^2}\bigl(C_{ij} T^\kappa_i-C_{ij}^0 T^0_i \bigr),
\nonumber\\
{d\sigma_R^F \over d\Gamma_0}&=&
{8 \pi\alpha^4 \over St^2}
\sum\limits_{i,j}
{\cal F}_j
\int\limits_0^{v^2_m} {dV^2} C_{ij} T^F_i .
\nonumber
\end{eqnarray}
Combining all contribution together we obtain the total BH cross section with the lowest order RC
\begin{equation}\label{fincs}
{d\sigma \over d\Gamma_0}=
\Bigl(1+{\alpha \over \pi}\delta\Bigr){d\sigma_0 \over d\Gamma_0}
+{d\sigma_v^F \over d\Gamma_0}
+{d\sigma_d \over d\Gamma_0}
+{d\sigma_R^F \over d\Gamma_0},
\end{equation}
where $\delta=\delta_{VR}+\delta_{vac}$ and
\begin{equation}
\delta_{VR}=
\frac{3}{2}L_m
-\frac{1}{2}
+\log{V_m^4\over u_0w_0}(L_m-1)
-\frac{1}{2}\log^2{u_0\over w_0}.
\end{equation}
This expression does not contain the infrared divergence, but also does not have terms with $L_m^2$. 

Combining leading terms allows us to reconstruct the results in the leading approximation obtained in \cite{AkushevichIlyichev2012} from the cross section (\ref{fincs}). Only the terms  $d\sigma_d / d\Gamma_0$, $d\sigma_R^F / d\Gamma_0$ and the correction factor $\delta$ contain the leading logs. The leading log terms in $d\sigma_d / d\Gamma_0$ and $d\sigma_R^F / d\Gamma_0$ are combined as:
\begin{eqnarray}
&&\sum_i C_{ij} \bigl({T^F_i \over 4\pi}+L_m{2T^\kappa_i \over V^2}\bigr)=
\nonumber
\\&&\qquad\qquad
{L_m\over 2}\biggl( 
{1+z_1^2\over z_1(1-z_1)}{{\hat T}^{0s}_j \over w}
+
{1+z_2^2\over 1-z_2}{{\hat T}^{0p}_j \over u} \biggr),
\end{eqnarray}
where 
\begin{equation}\label{z1z2uw}
z_1=1-{V^2\over w}, \quad z_2={u\over u+V^2}
\end{equation}
and ${\hat T}^{0s}_j$ and ${\hat T}^{0p}_j$ are obtained from ${\hat T}^{0}_j=\sum_i C_{ij}^0 T^{0}_i$ by substitution of kinematical variables by $z_{1,2}$-dependent variables as in (\ref{LLz}). Then straightforward calculation using (\ref{z1z2uw}) and eqs (42) of ref. \cite{AkushevichIlyichev2012} gives 
\begin{equation}\label{withIsIp}
{d\sigma_d \over d\Gamma_0}
+{d\sigma_R^F \over d\Gamma_0}=
{d\sigma_{LL} \over d\Gamma_0}+
{\alpha L_m \over 2 \pi}(I_s+I_p){d\sigma_{0} \over d\Gamma_0},
\end{equation}
where $d\sigma_{LL} / d \Gamma_0$ exactly represents the second and third terms (i.e., terms with integrals) of the leading log contribution given in eq. (48) of ref. \cite{AkushevichIlyichev2012} and $I_{s,p}$ are
\begin{eqnarray}
&&I_s=\int\limits_0^1dz_1\biggl({1+z_1^2\over 1-z_1}-{\sqrt{\lambda_Y}\over \sqrt{\lambda_{Yz}}}{\sin \theta' \over \sqrt{D_0}}{2\theta(z_1-z_1^m)\over 1-z_1}\biggr),\nonumber\\
&&I_p=\int\limits_0^1dz_2\biggl({1+z_2^2\over 1-z_2}-{\sqrt{\lambda_Y}\over \sqrt{\lambda_{Yz}}}{\sin \theta' \over \sqrt{D_0}}{2\theta(z_1-z_1^m)\over z_2(1-z_2)}\biggr).\nonumber
\end{eqnarray}
These integrals are calculated by a regularization at the upper integration limit as $1\rightarrow 1-\epsilon$. Then two terms can be calculated separately, e.g., 
\begin{equation}
\int\limits_0^{1-\epsilon} dz_1{1+z_1^2\over 1-z_1}=-{3\over 2}-2\log \epsilon
\end{equation}
and
\begin{equation}
\int\limits_{z_1^m}^{1-\epsilon}dz_1{\sqrt{\lambda_Y}\over \sqrt{\lambda_{Yz}}}{\sin \theta' \over \sqrt{D_0}}{2\over 1-z_1}
=
2\int\limits_{\epsilon w_0}^{V_m^2}{dV^2\over V^2}=2\log{V_m^2 \over \epsilon w_0}. 
\end{equation}
Therefore 
\begin{equation}
I_s+I_p=-3-2\log{V_m^4 \over  w_0 u_0}.
\end{equation}
One can see that the term with $I_s+I_p$ in (\ref{withIsIp}) exactly cancel the leading contribution from the correction factor $\delta$, such that the results (48) in \cite{AkushevichIlyichev2012} are reproduced.


\section{\label{na321}Numerical analysis}

Detailed numerical analysis of the leading log RC to the cross sections and spin asymmetries in the kinematics of JLab experiments was presented in ref. \cite{AkushevichIlyichev2012}. Here, we focus on the comparison of the effects at the leading and next-to-leading level and on some other effects not sufficiently discussed in that paper (e.g., $x$ and $Q^2$-dependence of RC).  

\subsection{RC for the cross section and polarization asymmetry} 

\begin{figure*}[htb]\centering
\scalebox{0.75}{\includegraphics{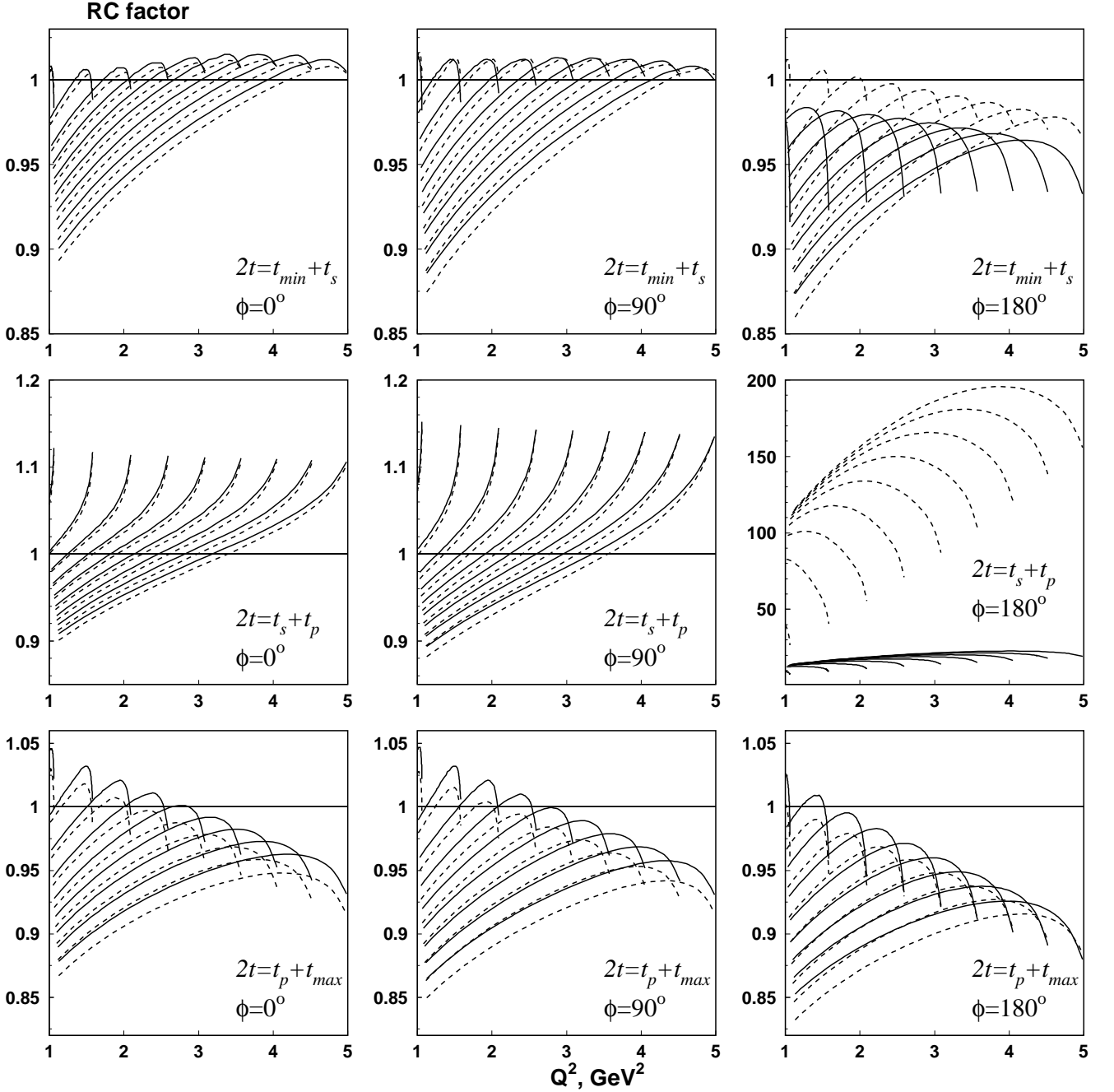}}
\hspace{0.4cm}
\caption{\label{xq2dep} The $Q^2$ dependence of LO (dashed) and NLO (solid) RC factor for several $x$, $t$ and $\phi$ calculated for beam energy equaling 5.75 GeV and without any cuts on the the invariant mass of two photons (i.e., when $V^2_m$ is given by eq. (\protect\ref{vmaxkin}). Nine curves at each plot correspond (from the left to the right) to nine values of $x$: 0.1, 0.15, 0.2, 0.25, 0.3, 0.35, 0.4, 0.45, and 0.5. $Q^2_{max}$ is defined by kinematics.}
\end{figure*}

Figure \ref{xq2dep} shows the $x$- and $Q^2$-dependencies of the RC factor (defined as $\delta_{RC}=\sigma_{obs}/\sigma_{BH}$, where $\sigma_{obs}$ is approximated by $d\sigma / d\Gamma_0$ from (\ref{fincs}) and $\sigma_{BH}=d\sigma_0 / d\Gamma_0$ is given by (\ref{sigmaBH001})) calculated in nine different kinematical points on $t$ and $\phi$. Three values of $\phi$ are 0$^o$, 90$^o$, and 180$^o$. Three points on $t$ represent three kinematical regions defining the shape of $t$-dependence of the RC factor (see, Figure 5 of ref. \cite{AkushevichIlyichev2012}). These regions are separated by the values of $t$ defining the position of the peaks in the cross section of BH process. The peaks correspond to emissions of the photon along the initial ($t_s=Q^2X/(S-Q^2)$) or final ($t_p=Q^2S/(X+Q^2)$) electrons. Specifically, these points are chosen as:
\begin{equation}
{t_{min}+t_s \over 2}, \quad
{t_{s}+t_p \over 2}, \quad
{t_{p}+t_{max} \over 2}. 
\end{equation}
Each plot contains eighteen curves (nine dashed curves representing LO calculation, nine solid curves representing NLO calculation) for nine $x$'s. The regions for $Q^2$ are between 1GeV$^2$ and $Q^2_{max}$ defined by kinematics:  $Q^2_{max}=xS^2/(S+xM^2)$. The shapes of the $Q^2$-dependencies are similar for LO and NLO cases, and the values of the NLO RC factors are typically higher than the RC factors calculated in the leading approximation; the curved for $\phi=$180$^o$ represent important exceptions that are discussed below. In many cases the NLO correction calculated in this paper is the significant contribution to the RC factor that should not be ignored in analyses of experimental data. Three types of the shapes of LO and NLO corrections are observed. The first is observed for upper and lower rows of the plots: the RC factor has a maximum for $Q^2$ close to $Q^2_{max}$. The monotonic curves are observed for  $\phi=$0$^o$ or 90$^o$ and for $t$ between two peaks (i.e., between $t_s$ and $t_p$). Third 
type 
of the shape is represented in the sixth plot of Figure \ref{xq2dep} (i.e., for $\phi=$180$^o$ and $t=(t_s+t_p)/2$). The large values of the RC factor found for this point deserve a specific attention that will be analyzed below in  Subsection \ref{largeeffect}.

\begin{figure}[htb]\centering
\scalebox{0.75}{\includegraphics{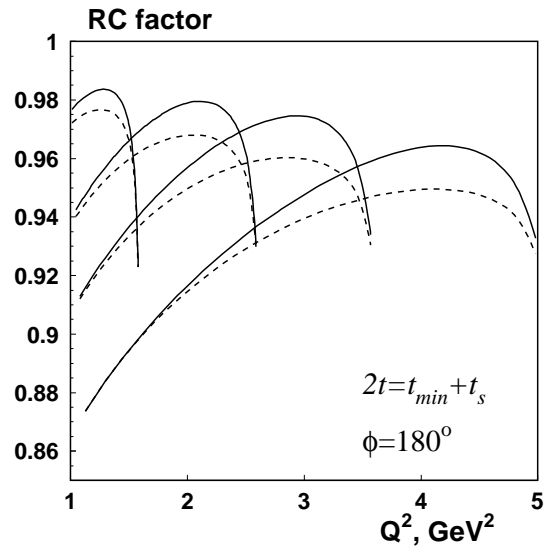}}
\hspace{0.4cm}
\caption{\label{cuts}  
The $Q^2$ dependence of NLO RC with (dashed) and without (solid) the cut on $V^2\leq 0.3$ GeV$^2$ several $x$, $t=(t_{min}+t_s)/2$, $\phi=$180$^o$ and $E_{beam}=$5.75 GeV. Four curves correspond (from the left to the right) to four values of $x$:  0.15, 0.25, 0.35, and 0.5. $Q^2_{max}$ is defined by kinematics.}
\end{figure}

Measurements of the  momenta of the final particles provide redundant information for the reconstruction of the four kinematical variables  determining the BH event. This information is used to construct various experimental cuts and apply them to data collected in a bin over these four variables. Largely these cuts are designed to remove the events with two photons (both of them are not soft) and could be reduced to the cut on $V^2$. These cuts remove events with $V^2>V^2_{cut}$. the cross section of the removed events are always positive therefore the effect of the cut results in decreasing the values of RC factor. The example of RC factors with and without a cut is presented in Figure 5 of ref. \cite{AkushevichIlyichev2012}. The value of decline in RC resulting from the cut on $V^2$ strongly depends on how this cut is hard. Figure \ref{cuts} gives an example how the cut on missing mass can influence RC. Since the cut can suppress only the contribution of two photon emission and respective 
cross section is always positive, the RC factor goes down when applying a harder cut on missing mass.  

\begin{figure}[htb]
\scalebox{0.55}{\includegraphics{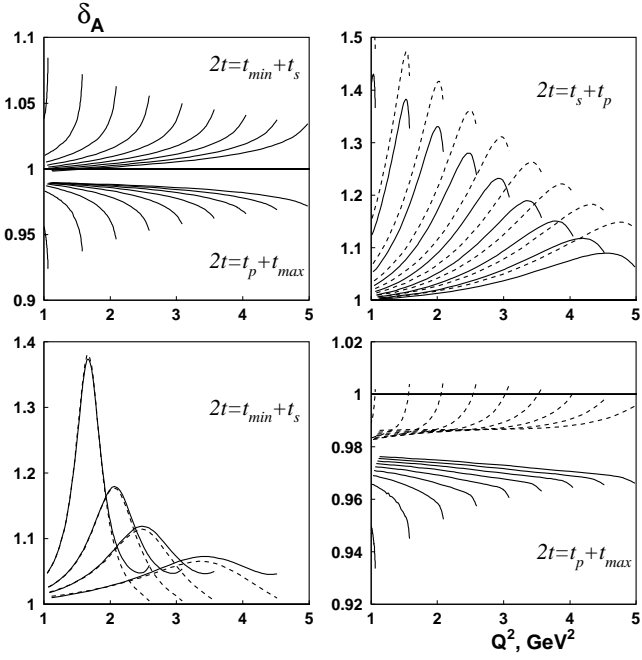}}
\hspace{0.4cm}
\caption{\label{xq2deppol} 
The $Q^2$ dependence of LO (dashed) and NLO (solid) radiative correction factors for polarization asymmetries (${\protect\vec \protect\eta}=(0,0,1)$ for upper plots, ${\protect\vec \protect\eta}=(1,0,0)$ for lower-left, and ${\protect\vec \protect\eta}=(0,1,0)$ for lower-right plots)
 for several $x$, $t$ and $\protect\phi=$180$^o$, $E_{beam}=$5.75 GeV without cuts on $V^2$. LO RC in the upper-left plot is not shown because $\protect\delta_A\approx 1$ for both values $t$. Nine curves at each plot correspond (from the left to the right) to nine values of $x$: 0.1, 0.15, 0.2, 0.25, 0.3, 0.35, 0.4, 0.45, and 0.5. $Q^2_{max}$ is defined by kinematics. Only curves for $x$: 0.25, 0.3, 0.35, and 0.5 are shown for lower-left plot. 
}
\end{figure}

The radiative correction factor can be defined for the polarization asymmetries in terms of RC factors for unpolarized and polarized parts of the cross section as 
$\delta_A=\delta_{RC}^{pol} / \delta_{RC}^{unp} $ because 
\begin{equation}
A_{obs}={\sigma_{obs}^{pol} \over \sigma_{obs}^{unp}}=
{\delta_{RC}^{pol} \over \delta_{RC}^{unp}} A_{BH}=
\delta_{A} A_{BH}.
\end{equation}
The index $obs$ and $BH$ stand for observed and BH cross section, superindices $unp$ and $pol$ characterize the polarization structure of the cross sections: 
\begin{equation}
\sigma_{obs,BH}=\sigma_{obs,BH}^{unp}+P_LP_N \sigma_{obs,BH}^{unp}.
\end{equation}
The radiative correction factor for asymmetries is not defined for kinematical points where $\sigma_{BH}^{pol}=0$. Several examples of $\delta_{A}$ are presented in Figure \ref{xq2deppol}. The RC factor to asymmetries is close to 1 comparing to RC factor for the unpolarized cross section. NLO correction could essentially change both magnitude and the shape of $\delta_{A}$. 

\subsection{Large effect for RC for $\phi=$180$^o$ and $t_s<t<t_p$} \label{largeeffect}

The nature of the large effect for $\phi=$180$^o$ and $t_s<t<t_p$ was clarified in \cite{AkushevichIlyichev2012}. The large effect comes from the two-photon
emission process when both two irradiated photons are
collinear: one is collinear to the initial electron and and another is collinear to the final electron. The corresponding
BH process (i.e., one photon emission process with the same values of the four kinematical variables) is the
process with the emitted photon with 4-momentum corresponding
to the sum of momenta of the two collinear
photons. This photon is not collinear and therefore the
respective cross section of the BH process is not large. This is visible from the leading log formulas (\ref{LmunuLL}). The first term in (\ref{LmunuLL}) corresponds to the $s$-peak of one of the photon and the second term describes the $p$-peak. If another photon is collinear to another electron (i.e., final electron for the first term and initial electron for the second term), respective scalar products of the photon with electron momenta (i.e., $u$ for the first term and $w$ for the second) has to be small. These scalar products appear in the denominator of the BH cross section making it large in this kinematical point.  

The origin of the effect  is visible from the NLO formulas as well. We see from (\ref{LLz}) that $u$ and $w$ in these terms both proportional to $V_{\mathrm I}$:
\begin{equation}
u={V_{\mathrm I} \over w}, \qquad w={V_{\mathrm I} \over u}. 
\end{equation}
Numerical analysis confirms that the huge effect appear due to the peak in the integrand in the point when $V_{\mathrm I}$ is small. 

This observation allows us to identify kinematical conditions when the effect can occur. Because $V_{\mathrm I}$ are proportional to the electron propagators we can expect that 
\begin{equation}\label{VIeq0}
V_{\mathrm I}=uw-V^2Q^2=0
\end{equation}
for $m\rightarrow 0$. Since $u$ and $w$ depend on $V^2$ (see eqs. (\ref{uwexplicit}) and \ref{lambdav})), eq. (\ref{VIeq0}) gives us an equation for finding $V^2$ representing the position of the peak of the integrand:
\begin{equation}\label{eqvi}
T_0^2+\sin^2(\phi)T_1T_2+\sin^4(\phi)^4T_1^2=0,
\end{equation}
where
\begin{eqnarray}
T_0&=&
4\lambda_Y(X+Q^2)(S-Q^2)
\biggl((t-t_p)(t-t_s)
\nonumber \\&&
+{X^2t_p t+S^2t_s t+2t_s t_pSX\over SXQ^2}V^2+{t_pt_s\over Q^4}V^4\biggr), 
\nonumber \\[1mm]
T_1&=&16Q^2(V^2_m-V^2)(SX-M^2Q^2)(S_xt
\nonumber \\[1mm]&&
+M^2(V^2+V^2_m+2t+2Q^2)),
\nonumber \\[1mm]
T_2&=&4\lambda_Y((S^2+X^2)((t-Q^2+V^2)^2+4Q^2V^2)
\nonumber \\[1mm]&&
+2Q^2t(S_{xt}(S_x-Q^2)-S_xV^2)).
\end{eqnarray}
The equation (\ref{eqvi}) was obtained by squaring of an equations containing radicals in the LHS or RHS. Therefore, some additional conditions have to be satisfied when searching for solutions of (\ref{eqvi}). One of them is 
\begin{equation}
\cos(\phi)<0.
\end{equation}

\begin{figure}[htb]\centering
\scalebox{0.75}{\includegraphics{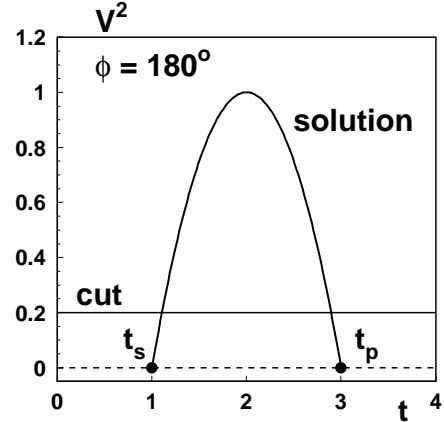}}
\hspace{0.4cm}
\caption{\label{solutionV2} The solution of the equation (\protect\ref{eqvi}) that is possible only $\protect\phi=180^o$ and $t_s \protect\leq t \protect\leq t_p$. 
The line at $V^2=0.2$ GeV$^2$ shows a possible cut on missing mass. This demonstrates us that the effect around the kinematical points $t=t_s$ and $t=t_p$ cannot be completely suppressed by the cut. }
\end{figure}

All three terms in (\ref{eqvi}) are non-negative (last term in $T_2$ is  positive because $S_{xt}(S_x-Q^2)-S_xV^2 \geq S_{xt}(S_x-Q^2)-S_xV^2_m=(S_x^2-Q^2t)^2/(tS_x^2)M^2+o(M^2/S_x)$), therefore the solution is possible only when all of them equal zero. This can be achieved only when $\phi=180^o$ and $T_0=0$. The solution of the latter equation is possible only for $t_s \leq t \leq t_p$ (otherwise all three terms in the brackets of the expression for $T_0$ are positive). The solution is $V^2=0$ for $t=t_{s,p}$ and some positive value of $V^2$ for $t$ between $t_s$ and $t_p$ as illustrated in Figure \ref{solutionV2}. Thus, we see that the extremely small values of $V_{\mathrm I}=0$ in the denominators of the integrand resulted in the huge RC can appear only for $t_s \leq t \leq t_p$  and   
$\phi=180^o$ (and a narrow area around the value). 

\section{\label{DC}Discussion and Conclusion}

In this paper we completed  analytic calculation of the radiative corrections in the BH process for unpolarized, 
longitudinally polarized, and transversely polarized targets within the next-to-leading accuracy. In this approximation only terms proportional the electron mass are neglected therefore the RC was calculated practically ``exactly''. The most important feature of the calculation is that 
complete integration for loop contributions and integration over 
angles of an additional photon for two-gamma contribution was performed analytically.  

This calculation is continuation of efforts of several groups of researches made contributions to the problem of RC in BH process which created a theoretical  background for our study. The one-loop correction and soft photon emission contributions were investigated in details by Vanderhaeghen et al. \cite{Vanderhaeghen2000}. Some ideas of one-loop correction calculation including ultraviolet and infrared renormalization using dimensional regularization were used in our calculation. The approach for the calculation in leading approximation for the DVCS cross section was developed by Bytev, Kuraev, and Tomasi-Gustafsson \cite{ByKuTo2008PRC}. We largely based on their approach for the calculation in the leading log approximation as well as compared our expression for the shifted kinematics and the expression for loops effect in the leading approximation. The explicit results in the leading log approximation which were used for the comparison were calculated by us in ref. \cite{AkushevichIlyichev2012}.
Two important sources of theoretical information were used for the NLO calculations. The first is the exact calculations of the next order corrections to the radiative tails from elastic peaks was done by Akhundov, Bardin, and Shumeiko \cite{AkBaSh1986YP}. Specifically, we used the model for the phase space parametrization of two photons and exact approach for extraction of infrared divergence (see also \cite{ABMO1985}). The second is the asymptotic expressions for loop integrals in non--collinear kinematics calculated for one-loop diagrams by Arbuzov, Belitsky, Kuraev, Shaikhatdenov \cite{Kuraevintegrals}. All these integrals were recalculated by us in dimensional regularization and the analytical expressions were tested numerically. Finally, we used the theory of DVCS from Belitsky, Mueller, Kirchner \cite{BMK2002} for cross check of the BH formulas for unpolarized and spin-dependent cross sections.

Leading log approximation provided compact expressions are a particular case of our calculation. We demonstrated how the leading log formulas can be extracted analytically and evaluated the relative contributions of the leading log and next-to-leading terms to the total RC. Note, that hadronic structure is incorporated in the leading log expressions through the ``base'' cross section (i.e., the cross section with one photon emission) that minimize  possible model-dependence in the RC.  

Large effects are predicted when both photons are collinear (one is collinear to the initial lepton and another is collinear to the final lepton). 
Since the photon in respective BH process is not collinear (its momentum is the sum of two collinear photons), the BH cross section is not so large. We found analytical expressions for kinematical conditions of the large effects and demonstrated that frequently used kinematical cuts cannot suppress this large effect especially in the kinematical region close to $t=t_s$ and $t=t_p$. 

All conclusions are valid for the specific way of reconstruction of kinematic variables: leptonic and hadronic momenta are used to reconstruct the kinematics of the BH process. Kinematical variables of the photon were assumed to be unmeasured (or used only in the calculation of kinematical cuts). An
universal way to avoid multiple calculations to cover all possibilities for data analysis designs is the development of the Monte Carlo generator of the BH process with the additional process with two photons.

\appendix

  \section{Loop Integrals  }\label{loopintegrals}
  
  In this Appendix we present the table of integrals over 4-momentum of an
  additional virtual photon that is denoted as $\ell$. The integrals are defined as:
  \begin{equation}
  J[A]={1\over i\pi^2} \int d^4 \ell A.
  \end{equation}
  
   The scalar, vector, and tensor integrals are denoted as $J_{ijk}$, $J^\mu_{ijk}$ and $J^{\mu\nu}_{ijk}$, respectively, where lower indices show the terms in integrand denominator, i.e., $i,j,k$ are 0, 1, 2, $q$, or $\bar q$ for $v_0$, $v_1$, $v_2$, $v_1^k$, or $v_2^k$, respectively, where
\begin{eqnarray}
   v_0&=&\ell^2, \nonumber\\ 
   v_1&=&(k_1-\ell)^2-m^2, \nonumber\\
   v_2&=&(k_2-\ell)^2-m^2, \nonumber\\
   v_1^k&=&(k_1-k-\ell)^2-m^2, \nonumber\\
   w_2^k&=&(k_2+k-\ell)^2-m^2. \nonumber
\end{eqnarray}
   For example, $J\biggl[{1\over v_0 v_1^k} \biggr]=J_{0q}$, $J[\frac{\ell_\mu}{v_0v_2v_2^k}]=J^\mu_{02\bar q}$, and $J[\frac{\ell_\mu\ell_\nu}{v_0v_1v_2v_2^k}]=J^{\mu\nu}_{012\bar q}$. 

The integration involves the usage of the Feynman parameters and is performed in the dimensional regularization. Note, that all expressions in this Appendix contain $w_0$ and $u_0$. However we drop the index "0" in this Appendix for brevity and use $w$ and $u$ instead. The results of the integration for scalar integrals are:   
  \begin{eqnarray}
   J_{01} &=& -2P+2,    \nonumber\\
   J_{02} &=& -2P+2, \nonumber\\
   J_{0q} &=&  -2P+2-L_w, \nonumber\\
   J_{0\bar q} &=&-2P+2-L_u, \nonumber\\
   J_{12} &=&  -2P+2-L_m, \nonumber\\
   J_{1q} &=& -2P, \nonumber\\
   J_{2\bar q} &=& -2P, \nonumber\\
   J_{2q} &=&  -2P+2-L_t, \nonumber\\
   J_{1\bar q} &=& -2P+2-L_t, \nonumber\\
    J_{012}  &=& -2\frac{L_m}{Q^2}P_{IR}-{1\over 2Q^2}\bigl( L_m^2-\frac{\pi^2}{3} \bigr), \nonumber\\
   J_{01q} &=&-\frac{1}{w}\bigl(\frac{1}{2}L_w^2+\frac{\pi^2}{3}\bigr), \nonumber\\
   J_{01\bar q}&=&-\frac{1}{Q^2+w}\bigl(L_t (L_t-L_u)
   \nonumber\\&&   \!\!\!\!\!\!+\frac{1}{2} (L_t-L_u)^2-\frac{\pi^2}{2}+2\Phi\bigl(1+\frac{u}{t}\bigr)\bigr),
   \nonumber\\
   J_{02q} &=&-\frac{1}{Q^2-u}\bigl(L_t (L_t-L_w)
   \nonumber\\&& +\frac{1}{2}(L_t-L_w)^2+2\Phi\bigl(1-\frac{w}{t}\bigr)\bigr),
   \nonumber\\
   J_{02\bar q}&=&\frac{1}{u}\bigl(\frac{1}{2}L_u^2-\frac{\pi^2}{6}\bigr), \nonumber\\
   J_{12q}&=&\frac{1}{2(u-w)}\bigl(L_t^2-L_m^2\bigr), \nonumber\\
   J_{12\bar q} &=&\frac{1}{2(u-w)}\bigl(L_t^2-L_m^2\bigr), 
   \nonumber\\
   J_{012q}&=&{2P_{IR}L_m\over wQ^2}+{1\over wQ^2}\biggl( 2L_mL_w-L_t^2
   \nonumber\\&& \qquad\qquad
   -2\Phi\Bigl(1-\frac{t}{Q^2}\Bigr)-\frac{\pi^2}{6}\biggr),
   \nonumber\\
   J_{012\bar q}&=&-{2P_{IR}L_m\over uQ^2}-{1\over uQ^2}\biggl( 2L_mL_u-L_t^2
   \nonumber\\&& \qquad\qquad
   -2\Phi\Bigl(1-\frac{t}{Q^2}\Bigr)-\frac{\pi^2}{6}\biggr).
\end{eqnarray}
Here 
$L_w=\log(w/m^2)$,
$L_u=\log(u/m^2)$,
$L_t=\log(t/m^2)$, and $L_m$ is defined after eq. (\ref{dirdm}).
The two-propagator vector integrals are 
\begin{eqnarray}
   \nonumber
   J_{01}^\mu &=& \left(-P+\frac{1}{2}\right)k_{1\mu}, \\
   J_{02}^\mu &=& \left(-P+\frac{1}{2}\right)k_{2\mu}, \nonumber\\
   J_{0q}^\mu &=& \left(-P+1-\frac{1}{2}L_w \right)(k_{1\mu}-k_{\mu}), \nonumber\\
   J_{0\bar q}^\mu  &=& \left(-P+1-\frac{1}{2}L_u \right)(k_{2\mu}+k_{\mu}), \nonumber\\
   J_{12}^\mu &=& \left(-P+1-\frac{1}{2}L_m \right)(k_{1\mu}+k_{2\mu}), \nonumber\\
   J_{1q}^\mu &=& -P(2k_{1\mu}-k_{\mu}), \nonumber\\
   J_{2\bar q}^\mu &=& -P(2k_{2\mu}+k_{\mu}), \nonumber\\
   J_{2q}^\mu  &=&  \left(-P+1-\frac{1}{2}L_t \right)(k_{1\mu}+k_{2\mu}-k_{\mu}), \nonumber\\
   J_{1\bar q}^\mu  &=& \left(-P+1-\frac{1}{2}L_t \right)(k_{1\mu}+k_{2\mu}+k_{\mu}).
\end{eqnarray}
 The result of vector and tensor integration for three-propagator integrals are
   \begin{eqnarray}
   \nonumber
   J^\mu_{ijk}&=&a_{ijk}^1 k_{1\mu}+a_{ijk}^2 k_{2\mu } - a_{ijk}^qq_\mu, 
\\
   J^{\mu\nu}_{ijk}&=&b_{ijk}^gg_{\mu\nu}+b_{ijk}^{11} k_{1\mu} k_{1\nu}
   +b_{ijk}^{22} k_{2\mu} k_{2\nu}
   +b_{ijk}^{qq} q_{\mu} q_{\nu}
\nonumber\\&&
   +b_{ijk}^{12} (k_{1\mu} k_{2\nu}+k_{2\mu} k_{1\nu})
   -b_{ijk}^{1q} (k_{1\mu} q_{\nu}+q_{\mu} k_{1\nu})
\nonumber\\&&
   -b_{ijk}^{2q} (k_{2\mu} q_{\nu}+q_{\mu} k_{2\nu}).
\end{eqnarray}
Integrals with four terms in denominators are defined in the same way but with four lower indices. The ``minus'' sign appeared in some terms because of opposite definition of $q$ used in \cite{Kuraevintegrals}. The coefficients for three-propagator vector integrals are 
\begin{eqnarray}
   && a_{012}^1=a_{012}^2={L_m\over -Q^2},\quad a_{012}^q=0, \nonumber\\
   && a_{01q}^1=\frac{1}{w}\Bigl(wJ_{01q}+2L_w-2\Bigr),\nonumber\\
   && a_{01q}^2=\frac{1}{w}\Bigl(-L_w+2\Bigr),\nonumber\\
   && a_{01q}^q=-\frac{1}{w}\Bigl(-L_w+2\Bigr),\nonumber\\
   && a_{01\bar q}^1=\frac{u}{w+Q^2}J_{01\bar q}-2\frac{t}{(w+Q^2)^2}L_t+\frac{t-u}{(w+Q^2)^2}L_u,\nonumber\\
   && a_{01\bar q}^2=0, \quad a_{01\bar q}^q=\frac{L_u-L_t}{w+Q^2},\nonumber\\
   && a_{02q}^1=0,  \quad a_{02q}^q=\frac{L_w-L_t}{u-Q^2},\nonumber\\
   && a_{02q}^2=\frac{w}{u-Q^2}J_{02q}-2\frac{t}{(u-Q^2)^2}L_t+\frac{t+w}{(u-Q^2)^2}L_w,\nonumber\\
   && a_{02\bar q}^1=a_{02\bar q}^q=\frac{1}{u}\Bigl(L_u-2\Bigr),\nonumber\\
   && a_{02\bar q}^2=\frac{1}{u}\Bigl(uJ_{02\bar q}-2L_u+2\Bigr),\nonumber\\
   && a_{12q}^1=\frac{t}{w-u}J_{12q}-\frac{t+Q^2}{(w-u)^2}L_m+2\frac{t}{(w-u)^2}L_t+\frac{2}{u-w},\nonumber\\
   && a_{12q}^2=J_{12q}-a_{12q}^1,\nonumber\\
   && a_{12q}^q=\frac{Q^2}{w-u}J_{12q}+\frac{t+Q^2}{(w-u)^2}L_t-2\frac{Q^2}{(w-u)^2}L_m+\frac{2}{u-w},\nonumber\\
   && a_{12\bar q}^2=\frac{t}{w-u}J_{12\bar q}-\frac{t+Q^2}{(w-u)^2}L_m+2\frac{t}{(w-u)^2}L_t+\frac{2}{u-w},\nonumber\\
   && a_{12\bar q}^1=J_{12\bar q}-a_{12\bar q}^2,\nonumber\\
   && a_{12\bar q}^q=\frac{Q^2}{u-w}J_{12\bar q}-\frac{t+Q^2}{(w-u)^2}L_t+2\frac{Q^2}{(w-u)^2}L_m\nonumber\\
&& \quad \quad \quad \quad -\frac{2}{u-w}.
\end{eqnarray}
   Ultraviolet divergent terms are contained in $b^g_{ijk}$ only:
\begin{eqnarray}
	 && b_{012}^{g}=-\frac{1}{2}P  +{3-L_m \over 4}, \nonumber\\
	 && b_{01q}^{g}=-\frac{1}{2}P  +{3-L_w \over 4},  \nonumber\\
	 && b_{01\bar q}^{g}=-\frac{1}{2}P -{uL_u+tL_t \over 4(w+Q^2)} +{3 \over 4}, \nonumber\\
	 && b_{02q}^{g}=-\frac{1}{2}P -{wL_w-tL_t \over 4(u-Q^2)} +{3 \over 4},    \nonumber\\
	 && b_{02\bar q}^{g}=-\frac{1}{2}P  +{3-L_u \over 4},  \nonumber\\
	 && b_{12q}^{g}=-\frac{1}{2}P  -{yL_m-tL_t \over 4(u-w)} +{3 \over 4},  \nonumber\\
	 && b_{12\bar q}^{g}=-\frac{1}{2}P -{yL_m-tL_t \over 4(u-w)} +{3 \over 4}. 
\end{eqnarray}
 The ultraviolet-free coefficients for three-propagator tensor integrals are   
\begin{eqnarray}
     && b_{012}^{11}=b_{012}^{22}=-{L_m-1\over 2Q^2}, \nonumber\\
     && b_{012}^{12}={-1\over 2Q^2}, \quad b_{012}^{1q}=b_{012}^{2q}=b_{012}^{qq}=0,\nonumber\\
	 && b_{01q}^{11}=J_{01q}+{3\over w}L_w-{9\over2w}, \nonumber\\
	 && b_{01q}^{12}=-{L_w-3\over 2w}, \nonumber\\
	 && b_{01q}^{1q}={L_w-3\over 2w}, \nonumber\\
	 && b_{01q}^{2q}=-b_{01q}^{22}=-b_{01q}^{qq}={L_w-2\over 2w}, \nonumber\\
	 && b_{01\bar q}^{11}=\biggl({u(t-u)\over (w+Q^2)^3}+{t^2+2tw-u^2\over 2(w+Q^2)^3}\biggr)L_u
	 \nonumber\\
	 &&\qquad
	 -{t(t+4u)\over (w+Q^2)^3}L_t+{u-t\over 2(w+Q^2)^2}+{u^2\over(w+Q^2)^2}J_{01\bar q}
	 , \nonumber\\
	 && b_{01\bar q}^{12}=b_{01\bar q}^{22}=b_{01\bar q}^{2q}=0, \nonumber\\
	 && b_{01\bar q}^{1q}={2u+t\over 2(w+Q^2)^2}(L_u-L_t)-{1\over2(w+Q^2)},\nonumber\\
	 && b_{01\bar q}^{qq}={L_u-L_t \over 2(w+Q^2)}, \nonumber\\
	 && b_{02q}^{22}=\biggl({w(t+w)\over (u-Q^2)^3}-{t^2-2tw-w^2\over 2(u-Q^2)^3}\biggr)L_w
	 \nonumber\\
	 &&\qquad
	 +{t(t-4w)\over (u-Q^2)^3}L_t-{w+t\over 2(u-Q^2)^2}+{w^2\over(u-Q^2)^2}J_{02q}
	 , \nonumber\\
	 && b_{02q}^{11}=b_{02q}^{12}=b_{02q}^{1q}=0, \nonumber\\
	 && b_{02q}^{2q}={2w-t\over 2(u-Q^2)^2}(L_w-L_t)-{1\over2(u-Q^2)}, \nonumber\\
	 && b_{02\bar q}^{22}=J_{02\bar q}-{3\over u}L_u+{9\over 2u}, \nonumber\\
	 && b_{02\bar q}^{12}=b_{02\bar q}^{2q}={L_u-3\over 2u}, \nonumber\\
	 && b_{02\bar q}^{11}=b_{02\bar q}^{qq}=b_{02\bar q}^{1q}={L_u-2\over 2u}, \nonumber\\
	 && b_{02\bar q}^{qq}={L_t-L_w\over 2(u-Q^2)}, \nonumber\\
	 && b_{12q}^{11}={t^2\over (u-w)^2}J_{12q}+{3t^2+4tQ^2-Q^4\over 2(u-w)^3}L_m
	 \nonumber\\&&\qquad\qquad
	 -{3t^2\over (u-w)^3}L_t
	 +{Q^2-4t\over (u-w)^2},
	 \nonumber\\
	 && b_{12q}^{22}={Q^4\over (u-w)^2}J_{12q}+{3Q^4+4tQ^2-t^2\over 2(u-w)^3}L_m
	 \nonumber\\&&\qquad\qquad
	 +{t(t-4Q^2)\over (u-w)^3}L_t
	 -{3Q^2\over (u-w)^2},
	 \nonumber\\
	 && b_{12q}^{qq}={Q^4\over (u-w)^2}J_{12q}+{3Q^4\over (u-w)^3}L_m
	 \nonumber\\&&\qquad\qquad
	 +{t^2-4tQ^2-3Q^4\over 2(u-w)^3}L_t
	 +{t-4Q^2\over (u-w)^2},
	 \nonumber\\
 	 && b_{12q}^{12}=-{tQ^2\over (u-w)^2}J_{12q}-{t^2+4tQ^2+Q^4\over 2(u-w)^3}L_m
	 \nonumber\\&&\qquad\qquad
	 +{t(t+2Q^2)\over (u-w)^3}L_t +{2t+Q^2\over (u-w)^2},
	 \nonumber\\
 	 && b_{12q}^{1q}={tQ^2\over (u-w)^2}J_{12q}+{Q^2(5t+Q^2)\over 2(u-w)^3}L_m
	 \nonumber\\&&\qquad\qquad
	 -{t(t+5Q^2)\over 2(u-w)^3}L_t -{3t+3Q^2\over 2(u-w)^2},
	 \nonumber\\
 	 && b_{12q}^{2q}=-{Q^4\over (u-w)^2}J_{12q}-{Q^2(t+5Q^2)\over 2(u-w)^3}L_m
	 \nonumber\\&&\qquad\qquad
	 +{2Q^4+5tQ^2-t^2\over 2(u-w)^3}L_t +{7Q^2-t\over 2(u-w)^2},
	 \nonumber\\
	 && b_{12\bar q}^{22}={t^2\over (u-w)^2}J_{12\bar q}+{3t^2+4tQ^2-Q^4\over 2(u-w)^3}L_m
	 \nonumber\\&&\qquad\qquad
	 -{3t^2\over (u-w)^3}L_t
	 +{Q^2-4t\over (u-w)^2},
	 \nonumber\\
	 && b_{12\bar q}^{11}={Q^4\over (u-w)^2}J_{12\bar q}+{3Q^4+4tQ^2-t^2\over 2(u-w)^3}L_m
	 \nonumber\\&&\qquad\qquad
	 +{t(t-4Q^2)\over (u-w)^3}L_t
	 -{3Q^2\over (u-w)^2},
	 \nonumber\\
	 && b_{12\bar q}^{qq}={Q^4\over (u-w)^2}J_{12\bar q}+{3Q^4\over (u-w)^3}L_m
	 \nonumber\\&&\qquad\qquad
	 +{t^2-4tQ^2-3Q^4\over 2(u-w)^3}L_t
	 +{t-4Q^2\over (u-w)^2},
	 \nonumber\\
 	 && b_{12\bar q}^{12}=-{tQ^2\over (u-w)^2}J_{12\bar q}-{t^2+4tQ^2+Q^4\over 2(u-w)^3}L_m
	 \nonumber\\&&\qquad\qquad
	 +{t(t+2Q^2)\over (u-w)^3}L_t +{2t+Q^2\over (u-w)^2},
	 \nonumber\\
 	 && b_{12\bar q}^{2q}=-{tQ^2\over (u-w)^2}J_{12\bar q}-{Q^2(5t+Q^2)\over 2(u-w)^3}L_m
	 \nonumber\\&&\qquad\qquad
	 +{t(t+5Q^2)\over 2(u-w)^3}L_t +{3t+3Q^2\over 2(u-w)^2},
	 \nonumber\\
 	 && b_{12\bar q}^{1q}={Q^4\over (u-w)^2}J_{12\bar q}+{Q^2(t+5Q^2)\over 2(u-w)^3}L_m
	 \nonumber\\&&\qquad\qquad
	 -{2Q^4+5tQ^2-t^2\over 2(u-w)^3}L_t -{7Q^2-t\over 2(u-w)^2}.
   \end{eqnarray}
 \begin{widetext}
The coefficients for four-propagator vector and tensor integrals are: 
 \begin{eqnarray}
   &&a_{012q}^1={1 \over 2uwQ^2}\biggl((wQ^2+ut)J_{12q}+(u-Q^2)^2J_{02q}
   -w(u+Q^2)J_{01q}-(u-Q^2)\Bigl({L_m^2\over 2}-2L_mL_w
   +L_t^2+2\Phi\Bigl(1-\frac{t}{Q^2}\Bigr)\Bigr)\biggl),
   \nonumber\\
   &&a_{012q}^2={1 \over 2uwQ^2}\biggl(-(uQ^2+wt)J_{12q}-(uw+tQ^2)J_{02q}
   +w(w+Q^2)J_{01q}-(w+Q^2)\Bigl({L_m^2\over 2}-2L_mL_w
   +L_t^2+2\Phi\Bigl(1-\frac{t}{Q^2}\Bigr)\Bigr)\biggl),
   \nonumber\\
      &&a_{012q}^q={1 \over 2uwQ^2}\biggl(Q^2(w+u)J_{12q}-Q^2(u-Q^2)J_{02q}
   -wQ^2J_{01q}+Q^2\Bigl({L_m^2\over 2}-2L_mL_w
   +L_t^2+2\Phi\Bigl(1-\frac{t}{Q^2}\Bigr)\Bigr)\biggl),
   \nonumber\\
   &&a_{012\bar q}^2={1 \over 2uwQ^2}\biggl(-(uQ^2+wt)J_{12\bar q}+(w+Q^2)^2J_{01\bar q}
   -u(w-Q^2)J_{02\bar q}+(w+Q^2)\Bigl({L_m^2\over 2}-2L_mL_u
   +L_t^2+2\Phi\Bigl(1-\frac{t}{Q^2}\Bigr)\Bigr)\biggl),
   \nonumber\\
   &&a_{012\bar q}^1={1 \over 2uwQ^2}\biggl((wQ^2+ut)J_{12\bar q}-(uw+tQ^2)J_{01\bar q}
   +u(u-Q^2)J_{02\bar q}+(u-Q^2)\Bigl({L_m^2\over 2}-2L_mL_u
   +L_t^2+2\Phi\Bigl(1-\frac{t}{Q^2}\Bigr)\Bigr)\biggl),
   \nonumber\\
      &&a_{012\bar q}^q=-{1 \over 2uwQ^2}\biggl(-Q^2(w+u)J_{12\bar q}+Q^2(w+Q^2)J_{01\bar q}
   +uQ^2J_{02\bar q}+Q^2\Bigl({L_m^2\over 2}-2L_mL_u
   +L_t^2+2\Phi\Bigl(1-\frac{t}{Q^2}\Bigr)\Bigr)\biggl),
   \nonumber\\
      &&b_{012q}^g={1 \over 2}\bigl(J_{12q}-wa_{012q}^q\bigr),
   \nonumber\\
   &&b_{012q}^{11}={1 \over 2uwQ^2}\biggl((u-Q^2)^2(J_{12q}-wa_{012q}^q)+(wQ^2+ut)a_{12q}^1
   -w(u+Q^2)a_{01q}^1+Q^2(u-Q^2)(a_{012}^1+wa_{012q}^1)\biggr),
   \nonumber\\
   &&b_{012q}^{22}={1 \over 2uwQ^2}\biggl((w+Q^2)^2(J_{12q}-wa_{012q}^q)-(uQ^2+wt)a_{12q}^2
   -(tQ^2+uw)a_{02q}^2+w(w+Q^2)a_{01q}^2
   \nonumber\\&&\qquad\qquad\qquad\qquad\qquad\qquad
   +Q^2(w+Q^2)(a_{012}^1+wa_{012q}^2)\biggr),
   \nonumber\\
   &&b_{012q}^{qq}={1 \over 2uwQ^2}\biggl(Q^4(J_{12q}-2wa_{012q}^q)+Q^2(u+w)a_{12q}^q
   +Q^2wa_{01q}^2-Q^2(u-Q^2)a_{02q}^q\biggr),   
   \nonumber\\
   &&b_{012q}^{12}={1 \over 2uwQ^2}\biggl(-(tQ^2+uw)(J_{12q}-wa_{012q}^q)-(uQ^2+wt)a_{12q}^1
   +w(w+Q^2)a_{01q}^1+Q^2(w+Q^2)(a_{012}^1+wa_{012q}^1)\biggr),
   \nonumber\\
   &&b_{012q}^{1q}={1 \over 2uwQ^2}\biggl(-Q^2(u-Q^2)(J_{12q}-2wa_{012q}^q)+(wQ^2+ut)a_{12q}^q
   +(u-Q^2)^2a_{02q}^q+w(u+Q^2)a_{01q}^2\biggr),
   \nonumber\\
   &&b_{012q}^{2q}={1 \over 2uwQ^2}\biggl(-Q^2(w+Q^2)(J_{12q}-2wa_{012q}^q)-(uQ^2+wt)a_{12q}^q
   -(Q^2t+uw)a_{02q}^q-w(w+Q^2)a_{01q}^2\biggr),
   \nonumber\\
      &&b_{012\bar q}^g={1 \over 2}\bigl(J_{12\bar q}-ua_{012\bar q}^q\bigr),
   \nonumber\\
   &&b_{012\bar q}^{22}={1 \over 2uwQ^2}\biggl((w+Q^2)^2(J_{12\bar q}-ua_{012\bar q}^q)-(uQ^2+wt)a_{12\bar q}^2
   -u(w-Q^2)a_{02\bar q}^2-Q^2(w+Q^2)(a_{012}^2-ua_{012\bar q}^2)\biggr),
   \nonumber\\
   &&b_{012\bar q}^{11}={1 \over 2uwQ^2}\biggl((u-Q^2)^2(J_{12\bar q}-ua_{012\bar q}^q)
   +(wQ^2+ut)a_{12\bar q}^1
   -(tQ^2+uw)a_{01\bar q}^1+u(u-Q^2)a_{02\bar q}^1
   \nonumber\\&&\qquad\qquad\qquad\qquad\qquad\qquad
   -Q^2(u-Q^2)(a_{012}^2-ua_{012\bar q}^1)\biggr),
   \nonumber\\
   &&b_{012\bar q}^{qq}={1 \over 2uwQ^2}\biggl(Q^4(J_{12\bar q}-2ua_{012q}^q)
   +Q^2(u+w)a_{12\bar q}^q
   -uQ^2a_{02\bar q}^1-Q^2(w+Q^2)a_{01\bar q}^q\biggr),   
   \nonumber\\
   &&b_{012\bar q}^{12}={1 \over 2uwQ^2}\biggl(-(tQ^2+uw)(J_{12\bar q}-ua_{012\bar q}^q)
   +(wQ^2+ut)a_{12q}^2
   +u(u-Q^2)a_{02\bar q}^2-Q^2(u-Q^2)(a_{012}^2-ua_{012\bar q}^2)\biggr),
   \nonumber\\
   &&b_{012\bar q}^{2q}=-{1 \over 2uwQ^2}\biggl(Q^2(w+Q^2)(J_{12\bar q}-2ua_{012\bar q}^q)+(uQ^2+wt)a_{12\bar q}^q
   -(w+Q^2)^2a_{01\bar q}^q+u(w-Q^2)a_{02\bar q}^1\biggr),
   \nonumber\\
   &&b_{012\bar q}^{1q}=-{1 \over 2uwQ^2}\biggl(Q^2(u-Q^2)(J_{12\bar q}-2ua_{012\bar q}^q)
   -(wQ^2+ut)a_{12\bar q}^q
   +(Q^2t+uw)a_{01\bar q}^q-u(u-Q^2)a_{02\bar q}^1\biggr).
   \nonumber
   \end{eqnarray}
  \end{widetext}
Note, that the integrals need to be considered in $d$-dimensional space. For
example, the results of two integrals $J_{012}^{\mu\nu}$ and $J_{12}$ gives an
equality $J_{012}^{\mu\nu}g_{\mu\nu}=J_{12}$ only with using dimensional
regularization rules, i.e., with taking into account that $P  g_{\mu\nu}
g_{\mu\nu}=Pd=4P+1+O(d-4)$. 
  
\section{Functions $T^v_{ij}$}\label{Vresults}

The functions $T^v_{ij}$ contributed to (\ref{ttvv}) read
\begin{eqnarray}
T^{v}_{10}&=&-{1\over u_0w_0}\bigl(4tt_y +3Q^2(t+Q^2)+ 8u_0w_0 \bigr),
\nonumber\\T^{v}_{11}&=&-{1\over  u_0^2w_0^2}\bigl(tQ^2t_y^2-(u_0^2+w_0^2+2Q^2)u_0w_0\bigr),
\nonumber\\T^{v}_{12}&=&{2\over u_0^2w_0}(u_0 (t_w^2+Q^4)-Q^2 t w_0),
\nonumber\\T^{v}_{13}&=&{2\over u_0w_0^2}(w_0 (t_u^2+Q^4)-Q^2 t u_0),
\nonumber\\T^{v}_{14}&=&{1\over u_0w_0} (4 (2 t_y t+u_0 w_0) Q^2 + 3 (u_0^2 +w_0^2) t_y), 
\nonumber\\T^{v}_{15}&=&{1\over u_0}(2 Q^2 + u_0) t_u, 
\nonumber\\T^{v}_{16}&=&{1\over u_0}(2 Q^2 - w_0) t_w, 
\nonumber\\T^{v}_{20}&=&{1\over 2u_0^2w_0^2}\bigl(3 t_y^3+4 (t Q^2+u_0 w_0) (5 t-Q^2)
+12 Q^2 u_0 w_0 \bigr),
\nonumber\\T^{v}_{21}&=&-{t\over 2u_0^3w_0^3}\bigl((t^2-Q^4)^2 + 2 u_0 w_0 (6 u_0 w_0+3 t^2+Q^4)\bigr),
\nonumber\\T^{v}_{22}&=&-{t\over u_0^3w_0}(4 t (Q^2+u_0)+6 u_0^2+t_y^2),
\nonumber\\T^{v}_{23}&=&-{t\over u_0w_0^3}(4 t (Q^2-w_0)+t_y^2+6 w_0^2),
\nonumber\\T^{v}_{24}&=&{1\over u_0^2w_0^2} (4 (2 t_y t+u_0 w_0) Q^2 + 3 (u_0^2 +w_0^2) t_y), 
\nonumber\\T^{v}_{25}&=&{t\over 2u_0^2w_0^2}(2 Q^2 + u_0) t_u, 
\nonumber\\T^{v}_{26}&=&{t\over 2u_0^2w_0^2}(2 Q^2 - w_0) t_w, 
\nonumber\\T^{v}_{30}&=&{1\over 2u_0^2w_0^2}\bigl(t_y^3-4 (Q^2-2 t) u_0 w_0 \bigr),
\nonumber\\T^{v}_{31}&=&-{t\over 2u_0^3w_0^3}\bigl(t_y^4+ 2 (u_0^2+w_0^2) u_0 w_0\bigr),
\nonumber\\T^{v}_{32}&=&-{t\over u_0^3w_0}(2 u_0^2+t_y^2),
\nonumber\\T^{v}_{33}&=&-{t\over u_0w_0^3}(2 w_0^2+t_y^2),
\nonumber\\T^{v}_{34}&=&{t\over u_0^2w_0^2} (t_y^2-4 u_0 w_0) t_y, 
\nonumber\\T^{v}_{35}&=&{t\over 2u_0^2w_0^2}((2 Q^2-3 u_0) t_y^2 - 2 u_0^2 (2 t-w_0)),
\nonumber\\T^{v}_{36}&=&{t\over 2u_0^2w_0^2}((2 Q^2+3 w_0) t_y^2 - 2 w_0^2 (2 t+u_0)), 
\nonumber\\T^{v}_{40}&=&-{u_0+w_0\over 2u_0^2w_0^2}\bigl(2 t (2 t+t_y)-3 t_y^2 \bigr),
\nonumber\\T^{v}_{41}&=&{t(u_0+w_0)\over 2u_0^3w_0^3}\bigl(2 t u_0 w_0+t_y^2 (t+Q^2)\bigr),
\nonumber\\T^{v}_{42}&=&{t\over u_0^3w_0}(2 Q^2 w_0+u_0^2+w_0^2),
\nonumber\\T^{v}_{43}&=&{t\over u_0w_0^3}(2 Q^2 u_0-u_0^2-w_0^2),
\nonumber\\T^{v}_{44}&=&{(u_0+w_0)\over u_0^2w_0^2} (2 u_0 w_0 Q^2-t_y^2 (t+Q^2)), 
\nonumber\\T^{v}_{45}&=&-{t\over 2u_0^2w_0^2}(2 w_0 Q^2 (2 t-w_0-u_0)-u_0 (3 w_0^2+u_0^2)), 
\nonumber\\T^{v}_{46}&=&{t\over 2u_0^2w_0^2}(-2 u_0 Q^2 (2 t+w_0+u_0)+w_0 (3 u_0^2+w_0^2)), 
\nonumber\\T^{v}_{50}&=&{(u_0+w_0)\over tu_0^2w_0^2}\bigl((t (3 Q^2-t) + u_0^2 + w_0^2) t - 3 t_y u_0 w_0 \bigr),
\nonumber\\T^{v}_{51}&=&-{(u_0+w_0)\over 2u_0^2w_0^2}\bigl((3 Q^2-t) t+u_0^2+w_0^2\bigr),
\nonumber\\T^{v}_{52}&=&-{1\over u_0^2w_0}((3 Q^2-t) t -2 u_0 t_y+u_0^2+w_0^2),
\nonumber\\T^{v}_{53}&=&-{1\over u_0w_0^2}((3 Q^2-t) t +2 w_0 t_y+u_0^2+w_0^2),
\nonumber\\T^{v}_{54}&=&-{u_0+w_0\over tu_0w_0} ((t+Q^2) t + 3 t_y^2),
\nonumber\\T^{v}_{55}&=&{1\over 2u_0w_0^2}(2 Q^2 (t Q^2+t_w^2)-(4 t_w-t_y) u_0 w_0), 
\nonumber\\T^{v}_{56}&=&{1\over 2u_0^2w_0}(2 Q^2 (t Q^2+t_u^2)-(4 t_u-t_y) u_0 w_0),
\nonumber\\T^{v}_{60}&=&-{1\over u_0^2w_0^2}\bigl(Q^2 t_y^2-(2 t+3 t_y) u_0 w_0 \bigr),
\nonumber\\T^{v}_{61}&=&{1\over 2u_0^2w_0^2}\bigl(t_y^2 Q^2 -2 u_0 w_0 (t+Q^2)\bigr),
\nonumber\\T^{v}_{62}&=&-{1\over u_0^2w_0}(2 u_0 t_w-t_y Q^2),
\nonumber\\T^{v}_{63}&=&-{1\over u_0w_0^2}(2 t_u w_0+t_y Q^2),
\nonumber\\T^{v}_{64}&=&-{1\over u_0w_0} (2 t+3 Q^2) t_y, 
\nonumber\\T^{v}_{65}&=&-{1\over 2u_0w_0^2}((2 Q^2 t-w_0 u_0) (w_0+u_0)-2 w_0 t_w t_y),
\nonumber\\T^{v}_{66}&=&{1\over 2u_0^2w_0}(-(2 Q^2 t-w_0 u_0) (w_0+u_0)+2 u_0 t_u t_y).
\end{eqnarray}
Notation used in these formulas are
\begin{equation}
t_y=t-Q^2, \quad 
t_w=t-w_0, \quad
t_u=t+u_0.
\end{equation}

\section{Integrals over phase space of two real photons }\label{Rintegrals}

  In this Appendix we present the table of two-dimensional integrals over $d\Gamma_{2\gamma}$. The integrals are defined as:
  \begin{equation}
  {\mathrm J}[A]={2\over \pi} \int d\Gamma_{2\gamma} A=
  {1\over 4\pi}\int d\cos\theta_R d\phi_R A.
  \end{equation}
There are two types of integrals to calculate:
\begin{equation}
J_{ij}={\mathrm J}\biggl[ {1 \over u_1^i w_2^j} \biggr], \quad 
{\bar J}_{ij}={\mathrm J}\biggl[ {1 \over u_1^i w_1^j} \biggr]
\end{equation}
The invariants $w_{1,2}$ and $u_{1,2}$ are defined after eq.(\ref{F12IR}). The integration is performed in the center-of-mass system of two photons. The axis $OZ$ is along $\bf q$ and the momenta of electrons are in the $OXZ$ plane. Photon energies are equal $\omega_1 =\omega_2 = {V /2}$. The electron energies and electron momenta and their the components in this system are 
\begin{eqnarray}
\nonumber
&&E_1={w \over 2V}, \quad E_2={u \over 2V}, \\
&&\quad p_{1z}={w^2-wu+2Q^2V^2 \over 2V \sqrt{\lambda_V}},
\nonumber\\
&&\quad p_{2z}={wu-u^2-2Q^2V^2 \over 2V \sqrt{\lambda_V}}, 
\nonumber\\
&& p_{1x}^2=p_{2x}^2=p_x^2={Q^2V_{\mathrm I} \over \lambda_V}-m^2, 
\end{eqnarray}
where $\lambda_V=(w-u)^2+4Q^2V^2$ 
and $V_{\mathrm I}$ is defined after eq.~(\ref{T123456}).
The invariants $w_{1,2}$ and $u_{1,2}$ are
\begin{eqnarray}
&&w_1 = 2\omega_1(E_1-p_{1z}\cos(\theta_R)-p_x\sin(\theta_R)\cos(\phi_R)), \nonumber\\
&&w_2=  2\omega_2(E_1+p_{1z}\cos(\theta_R)+p_x\sin(\theta_R)\cos(\phi_R)), \nonumber\\ 
&&u_1 = 2\omega_1(E_2-p_{2z}\cos(\theta_R)-p_x\sin(\theta_R)\cos(\phi_R)), \nonumber\\ 
&&u_2 = 2\omega_2(E_2+p_{2z}\cos(\theta_R)+p_x\sin(\theta_R)\cos(\phi_R)). \nonumber
\end{eqnarray}
The results of integration for scalar integrals with no more than two propagators are 
\begin{eqnarray}
&&J_{00}=1, \qquad 
J_{10}={L_u^V \over u}, \qquad
J_{01}={L_w^V \over w}, \nonumber\\
&&J_{20}={1\over m^2V^2}, \qquad
J_{02}={1\over m^2V^2}, \nonumber\\
&&{\bar J}_{11}= {2 L_m \over Q^2V^2}, \qquad
J_{11}= {2 L_{\mathrm I} \over V_{\mathrm I}}, \nonumber
\end{eqnarray}
where
\begin{eqnarray}
&&L_u^V=\log\frac{u^2}{m^2V^2}, \qquad
L_w^V=\log\frac{w^2}{m^2V^2},\nonumber \\
&&L_{\mathrm I}=\log\frac{V_{\mathrm I}}{m^2V^2}
\nonumber
\end{eqnarray}
and $L_m$ is defined after eq. (\ref{dirdm}).

The integration of three- and four-propagator integrals results in 
\begin{eqnarray}
J_{21}&=&{u\over m^2 V^2 V_{\mathrm I}} + {2 w \over V_{\mathrm I}^2} L_{\mathrm I} 
+ 2{ u -  w \over V_{\mathrm I}^2}, \nonumber\\
J_{12}&=&{w\over m^2 V^2 V_{\mathrm I}} + {2 u \over V_{\mathrm I}^2} L_{\mathrm I} - 2{u - w \over V_{\mathrm I}^2}, \nonumber\\
J_{22}&=&{u^2 + w^2 \over m^2 V^2 V_{\mathrm I}^2} - 4 {V_{\mathrm I} - 2 u w \over V_{\mathrm I}^3} L_{\mathrm I} 
+ 4 {u^2 + V_{\mathrm I} - 3 u w + w^2 \over V_{\mathrm I}^3}. \nonumber
\end{eqnarray}
The vector and tensor integrals are defined as
\begin{equation}
J_{ij}^\mu={\mathrm J}\biggl[ {\Delta_{\mu} \over u_1^i w_2^j} \biggr], \quad 
{\bar J}_{ij}^\mu={\mathrm J}\biggl[ {\Delta_{\mu} \over u_1^i w_1^j} \biggr],
\end{equation}
and 
\begin{equation}
J_{ij}^{\mu\nu}={\mathrm J}\biggl[ {\Delta_{\mu}\Delta_{\nu} \over u_1^i w_2^j} \biggr], \quad 
{\bar J}_{ij}^{\mu\nu}={\mathrm J}\biggl[ {\Delta_{\mu}\Delta_{\nu} \over u_1^i w_1^j} \biggr],
\end{equation}
with $\Delta=\kappa_{2}-\kappa_{1}$. The vector integrals with the dependence on one of $u_1$ or $w_2$ are
\begin{eqnarray}
 J^\mu_{10}&=&{L_u^V - 2 \over u} \bigl(\kappa_\mu - 2 {V^2\over u} k_{2\mu} \bigr),\nonumber\\
 J^\mu_{01}&=&{L_w^V - 2 \over w} \bigl(\kappa_\mu - 2 {V^2\over w} k_{1\mu} \bigr),\nonumber\\
 J^\mu_{20}&=&{1 \over m^2} \biggl( {\kappa_\mu\over V^2} - {2 k_{2\mu}\over u}\biggr) 
 + 2 {L_u^V - 2 \over u^2} \biggl({2 V^2\over u} k_{2\mu} - \kappa_{\mu}\biggr), \nonumber\\
 J^\mu_{02}&=&{1\over m^2}\biggl({\kappa_\mu\over V^2} - {2 k_{1\mu}\over w}\biggr) 
 + 2 {L_w^V - 2\over w^2} \biggl(2 {V^2\over w} k_{1\mu} - \kappa_\mu\biggr).\nonumber 
 \end{eqnarray}
  \begin{widetext}
  Remaining vector integrals and tensor integrals depending on one of $u_1$ or $w_2$ are
 \begin{eqnarray}
 {\bar J}^\mu_{11}&=&\biggl({2 u w L_m \over  V_{\mathrm I} Q^2 V^2} - {L_u^V + L_w^V\over 
    V_{\mathrm I}}\biggr) \kappa_\mu + \biggl(u {L_u^V - L_w^V - 2 L_m\over V_{\mathrm I} Q^2} 
+ 2 {V^2\over  V_{\mathrm I} w} L_w^V\biggr) k_{1\mu} 
	+ \biggl( { 2 L_u^V V^2\over u V_{\mathrm I}} 
+ {(-2 L_m - L_u^V + L_w^V) w\over Q^2 V_{\mathrm I}}\biggr) k_{2\mu} ,\nonumber\\
 J^\mu_{11}&=&{L_u^V - L_w^V\over 
  V_{\mathrm I}} \kappa_\mu + \biggl( {2 L_w^V V^2\over V_{\mathrm I} w} - {(-2 L_{\mathrm I} + L_u^V + L_w^V) u \over 
    Q^2 V_{\mathrm I} }\biggr) k_{1\mu} + \biggl({-2 L_u^V V^2\over u V_{\mathrm I}} + {(-2 L_{\mathrm I} + L_u^V + L_w^V) w\over 
    Q^2 V_{\mathrm I}}\biggr) k_{2\mu}, \nonumber\\
 J^\mu_{21}&=&\biggl({u\over m^2 V^2 V_{\mathrm I}} + 2 {u - (L_{\mathrm I} - 1) w\over V_{\mathrm I}^2}\biggr) \kappa_\mu + 
 4 {(L_{\mathrm I} - 1) V^2\over V_{\mathrm I}^2} k_{1\mu} - 2 \biggl({1 \over m^2 V_{\mathrm I}} + {2 V^2\over V_{\mathrm I}^2} \biggr) k_{2\mu}, \nonumber\\
 J^\mu_{12}&=&-\biggl({w\over m^2 V^2 V_{\mathrm I}} + 2 {w - (L_{\mathrm I} - 1) u\over V_{\mathrm I}^2}\biggr) \kappa_\mu - 
 4 {(L_{\mathrm I} - 1) V^2\over V_{\mathrm I}^2} k_{2\mu} + 2 \biggl({1 \over m^2 V_{\mathrm I}} + {2 V^2\over V_{\mathrm I}^2}\biggr) k_{1\mu}, \nonumber\\
 J^\mu_{22}&=&\biggl({1 \over m^2 V^2 V_{\mathrm I}^2} + {4 \over  V_{\mathrm I}^3}\biggr) (u^2 - w^2) \kappa_\mu + 
 2 \biggl({w\over m^2 V_{\mathrm I}^2} + {2 V^2 ((2 L_{\mathrm I} - 3) u + 2 w)\over  V_{\mathrm I}^3}\biggr) k_{1\mu} 
 \nonumber\\&&
 - 
 2 \biggl({u\over m^2 V_{\mathrm I}^2} + {2 V^2 ((2 L_{\mathrm I} - 3) w + 2 u)\over  V_{\mathrm I}^3}\biggr) k_{2\mu},
 \nonumber\\
 J^{\mu\nu}_{10}&=& {1\over u^3}\biggl(u^2 \bigl((L_u^V - 2) \kappa_\mu \kappa_\nu - V^2 g_{\mu\nu}\bigr) -  2 (L_u^V - 3) u V^2 (\kappa_\mu k_{2\nu} + k_{2\mu} \kappa_\nu) + 4 (L_u^V - 3) V^4 
 k_{2\mu} k_{2\nu}\biggr),
\nonumber\\
J^{\mu\nu}_{01}&=& {1\over w^3}\biggl( w^2\bigl((L_w^V-2) \kappa_\mu \kappa_\nu - V^2 g_{\mu\nu}\bigr) - 2 (L_w^V - 3) V^2 w (k_{1\mu} \kappa_\nu + \kappa_\mu k_{1\nu})
   + 4 (L_w^V - 3) V^4 k_{1\mu} k_{1\nu}\biggr), 
   \nonumber\\
J^{\mu\nu}_{20}&=& {1\over m^2V^2u^2}\biggl(u^2 \kappa_\mu \kappa_\nu - 2 u V^2 (\kappa_\mu k_{2\nu} + k_{2\mu} \kappa_\nu) + 4 V^4 k_{2\mu} k_{2\nu}\biggr)
 \nonumber\\&&
 + {1\over u^4}\biggl( u^2 (4 (3 - L_u^V) \kappa_\mu \kappa_\nu + 2 (2 - L_u^V) g_{\mu\nu} V^2) + 
    4 (3 L_u^V - 8) u V^2 (k_{2\mu} \kappa_\nu + \kappa_\mu k_{2\nu}) + 8 (8 - 3 L_u^V) V^4 k_{2\mu} k_{2\nu}\biggr),
	\nonumber\\
J^{\mu\nu}_{02}&=& {1\over m^2V^2w^2}\biggl(4 V^4 k_{1\mu} k_{1\nu} - 2 V^2 w (\kappa_\mu k_{1\nu} + k_{1\mu} \kappa_\nu) + w^2 \kappa_\mu \kappa_\nu \biggr) 
\nonumber\\&&
+  {1\over w^4}\biggl(
   w^2 (4 (3 - L_w^V) \kappa_\mu \kappa_\nu + 2 (2 - L_w^V) g_{\mu\nu} V^2) + 
    4 (3 L_w^V - 8) w V^2 (k_{1\mu} \kappa_\nu + \kappa_\nu k_{1\mu}) + 8 (8 - 3 L_w^V) V^4 k_{1\mu} k_{1\nu}\biggr).\nonumber
\end{eqnarray}
  \end{widetext}
Here
\begin{eqnarray}
L^y_{uw}&=&2L_m-L_w^V-L_u^V, \nonumber \\
{\bar L}_{uw}&=&2L_{\mathrm I}-L_w^V-L_u^V, \nonumber\\
 L_{uw}&=&2L_{\mathrm I}+L_u^V-L_w^V, \nonumber\\
 L_{wu}&=&2L_{\mathrm I}+L_w^V-L_u^V. 
\end{eqnarray}
Remaining integrals will be represented in the form
\begin{eqnarray}
   J^{\mu\nu}_{ij}&=&c_{ij}^gg_{\mu\nu}
   +c_{ij}^{11} k_{1\mu} k_{1\nu}
   +c_{ij}^{22} k_{2\mu} k_{2\nu}
   +c_{ij}^{kk} \kappa_{\mu} \kappa_{\nu}
   \nonumber\\&&
   +c_{ij}^{12} (k_{1\mu} k_{2\nu}+k_{2\mu} k_{1\nu})
   +c_{ij}^{1k} (k_{1\mu} \kappa_{\nu}+\kappa_{\mu} k_{1\nu})
   \nonumber\\&&
   +c_{ij}^{2k} (k_{2\mu} \kappa_{\nu}+\kappa_{\mu} k_{2\nu}).
\end{eqnarray}
The integrals  ${\bar J}^{\mu\nu}_{ij}$ have the same form and expressed in terms of coefficients ${\bar c}_{ij}^{\cdot}$. The coefficients of tensor integration are:
\begin{eqnarray}
\nonumber
{\bar c}_{11}^g &=& {L^y_{uw} V^2\over V_{\mathrm I}},
\\
{\bar c}_{11}^{11} &=&{
  2 V^2\over V_{\mathrm I}^2 w^2
    Q^2}
	\biggl(2 (L_w^V - 2) V_{\mathrm I}^2 + 2 u w V_{\mathrm I} + L^y_{uw} u^2 w^2\biggr),
\nonumber\\
{\bar c}_{11}^{22} &=&{2 V^2 \over 
  u^2 V_{\mathrm I}^2 Q^2}
\biggl(2 (L_u^V - 2) V_{\mathrm I}^2 + 2 u w V_{\mathrm I} + L^y_{uw} u^2 w^2\biggr),
\nonumber\\
{\bar c}_{11}^{kk} &=& {4\over V_{\mathrm I}} + {2 L^y_{uw} u w\over V_{\mathrm I}^2} + {2 L_m\over Q^2 V^2},
\nonumber\\
{\bar c}_{11}^{12} &=&{2 V^4 L^y_{uw} \over V_{\mathrm I}^2} + {4 V^2 \over Q^2 V_{\mathrm I}},
\nonumber\\
{\bar c}_{11}^{1k} &=& -{1\over 
   Q^2 V_{\mathrm I}^2 w}\biggl(2 (L_w^V - 2) V_{\mathrm I}^2 + (4 - L^y_{uw}) V_{\mathrm I} u w 
   \nonumber\\&&
   + 2 L^y_{uw} u^2 w^2\biggr),
\nonumber\\
{\bar c}_{11}^{2k} &=&{1\over   Q^2 V_{\mathrm I}^2 u}
\biggl(-2 (L_u^V - 2) V_{\mathrm I}^2 - (4 - L^y_{uw}) V_{\mathrm I} u w 
   \nonumber\\&&
- 2 L^y_{uw} u^2 w^2\biggr),
\nonumber\\
c_{11}^g &=& {{\bar L}_{uw} \over Q^2},
\nonumber\\
c_{11}^{11} &=& 
  {2\over V_{\mathrm I} w^2 Q^4} \biggl(2 (L_w^V - 2) V_{\mathrm I}^2 + 
     2 (3 - 2 L_w^V) u w V_{\mathrm I}  
        \nonumber\\&&
+ ( L_{wu}-2) u^2 w^2 \biggr),
\nonumber\\
c_{11}^{22} &=& {2 \over u^2 V_{\mathrm I} Q^4}\biggl(2 (L_u^V-2) V_{\mathrm I}^2 + 
     2 (3 - 2 L_u^V) u w V_{\mathrm I} 
        \nonumber\\&&
+ (L_{uw}-2) u^2 w^2\biggr),
\nonumber\\
c_{11}^{kk} &=& {2 (L_{\mathrm I} - 2)\over V_{\mathrm I}},
\nonumber\\
c_{11}^{12} &=& {2\over Q^4 V_{\mathrm I}} \biggl(-2 Q^2 V^2 - {\bar L}_{uw} V_{\mathrm I}\biggr),
\nonumber\\
c_{11}^{1k} &=& {1\over Q^2 V_{\mathrm I} w}\biggl((4 - L_{wu}) Q^2 V^2 - {\bar L}_{uw} V_{\mathrm I}\biggr),
\nonumber\\
c_{11}^{2k} &=&{1\over Q^2 u V_{\mathrm I}}\biggl((4 - L_{uw}) Q^2 V^2 - {\bar L}_{uw} V_{\mathrm I}\biggr),
\nonumber\\
c_{21}^{g} &=&{1\over Q^2 u V_{\mathrm I}}\bigl(2 L_u^V V_{\mathrm I} - L_{uw} u w\bigr),
\nonumber\\
c_{21}^{11} &=& {1\over   Q^4 V_{\mathrm I}^2 w}\bigl(
  4 L_w^V V_{\mathrm I}^2 + 2 u w (2 (1 - L_{wu}) V_{\mathrm I}
     \nonumber\\&&
+ (L_{wu} - 2) u w)\bigr),
\nonumber\\
c_{21}^{22} &=& {4 V^2\over m^2 u V_{\mathrm I}} + 
   {2\over Q^4 u^3 V_{\mathrm I}^2} \bigl(2 (2 - L_u^V) V_{\mathrm I}^2 (2  V_{\mathrm I} - 3 u w) 
      \nonumber\\&&
- u^3 w^3 (L_{uw} - 2) + 
      2 u^2 (2 Q^4 V^4 + w^2 V_{\mathrm I})\bigr),
\nonumber\\
c_{21}^{kk} &=& {u\over m^2 V^2 V_{\mathrm I}} + {1\over u V_{\mathrm I}^2}\bigl(
   4 V_{\mathrm I} + 2 u^2
      \nonumber\\&&
- 2 u (-1 + L_{\mathrm I} + L_u^V - L_w^V) w\bigr),
\nonumber\\
c_{21}^{12} &=& -{2\over    Q^4 u V_{\mathrm I}^2} \bigl(2 L_u^V V_{\mathrm I}^2 + 2 (1 - L_{uw}) V_{\mathrm I} u w
   \nonumber\\&&
+ (L_{uw}-2) u^2 w^2\bigr),
\nonumber\\
c_{21}^{1k} &=& {2 (L_u^V - L_w^V) V^2\over V_{\mathrm I}^2 },
\nonumber\\
c_{21}^{2k} &=&-{2\over m^2 V_{\mathrm I}} + {2\over    Q^2 u^2 V_{\mathrm I}^2}
 \bigl(-2 (u^2 + V_{\mathrm I}) u w 
    \nonumber\\&&
+ 
      2 V_{\mathrm I} (u^2 - (L_u^V - 2) V_{\mathrm I}) + ( L_{uw} - 2) u^2 w^2\bigr),
\nonumber\\
c_{12}^{g} &=& {1\over Q^2 V_{\mathrm I} w}\bigl(2 L_w^V V_{\mathrm I} - L_{wu}  u w\bigr),
\nonumber\\
c_{12}^{11} &=& {4 V^2\over m^2 V_{\mathrm I} w} + 
   {2\over Q^4 V_{\mathrm I}^2 w^3} \bigl(2 (2 - L_w^V) V_{\mathrm I}^2 (2  V_{\mathrm I} - 3 u w) 
      \nonumber\\&&
- u^3 w^3 (L_{wu} - 2) + 
      2 w^2 (2 Q^4 V^4 + u^2 V_{\mathrm I})\bigr),
\nonumber\\
c_{12}^{22} &=& {1\over   Q^4 u V_{\mathrm I}^2} \bigl(
  4 L_u^V V_{\mathrm I}^2 + 2 u w (-2 (L_{uw}-1) V_{\mathrm I}
     \nonumber\\&&
+ (L_{uw}-2) u w)\bigr),
\nonumber\\
c_{12}^{kk} &=&  {w\over m^2 V^2 V_{\mathrm I}} + {2\over w V_{\mathrm I}^2} \bigl(1 - L_{\mathrm I} + L_u^V - L_w^V) u w    \nonumber\\&&
+ 4 V_{\mathrm I} + 2 w^2\bigr),
   \nonumber\\
c_{12}^{12} &=& -{2\over 
   Q^4 V_{\mathrm I}^2 w} \bigl(2 L_w^V V_{\mathrm I}^2 + 
      2 (1 - L_{wu}) V_{\mathrm I} (Q^2 V^2 + V_{\mathrm I})   \nonumber\\&&
 + ( L_{wu}-2) u^2 w^2\bigr),
\nonumber\\
c_{12}^{1k} &=& -{2\over m^2 V_{\mathrm I}} - 
   {2\over Q^2 V_{\mathrm I}^2 w^2} \bigl(2 ( L_w^V-2) V_{\mathrm I}^2    \nonumber\\&&
+ 2 V_{\mathrm I} (u - w) w + 
      u w^2 ((2 - L_{wu}) u + 2 w)\bigr),
\nonumber\\
c_{12}^{2k} &=&{2 ( L_w^V-L_u^V) V^2\over V_{\mathrm I}^2},
\nonumber\\
c_{22}^{g} &=& -{4 (L_{\mathrm I}-1) V^2\over V_{\mathrm I}^2},
\nonumber\\
c_{22}^{11} &=& {4 V^2\over m^2 V_{\mathrm I}^2} + {16 V^4\over V_{\mathrm I}^3},
\nonumber\\
c_{22}^{22} &=& {4 V^2\over m^2 V_{\mathrm I}^2} + {16 V^4\over V_{\mathrm I}^3},
\nonumber\\
c_{22}^{kk} &=&{u^2 + w^2\over m^2 V^2 V_{\mathrm I}^2} + {4\over V_{\mathrm I}^3} \bigl(u^2 + (L_{\mathrm I} - 1) V_{\mathrm I} 
\nonumber\\&&
+ (3 - 2 L_{\mathrm I}) u w    
+ w^2\bigr),
\nonumber\\
c_{22}^{12} &=& {8 (3 - 2 L_{\mathrm I}) V^4\over V_{\mathrm I}^3},
\nonumber\\
c_{22}^{1k} &=& -{2 w\over m^2 V_{\mathrm I}^2} + {4 V^2\over V_{\mathrm I}^3} \bigl((2 L_{\mathrm I} - 3) u - 2 w\bigr),
\nonumber\\
c_{22}^{2k} &=& -{2 u\over m^2 V_{\mathrm I}^2} - {4 V^2\over V_{\mathrm I}^3} \bigl(2 u + (3 - 2 L_{\mathrm I}) w\bigr).
\end{eqnarray}

\section{Results for $T^F_{ij}$}\label{Rappendix}
  
The functions $T^F_{ij}$ contributed to (\ref{TiF}) read
\begin{widetext}  
\begin{eqnarray}\label{AppDform}
T^F_{11}&=&
6
+{V^2 + 2 w - u \over Q^2}
+{Q^2 \over w}
 - {u Q^2-Q^2_u (Q^2+V^2) \over Q^2 z_1}
 - w{ V^2 + w + 4 Q^2 \over Q^2 z_2}
+{Q^2_u V^2 \over z_1^2}
 - Q^2_{v2} t_y {Q^2 + V^2 \over Q^2 z_1 z_2}
 - 2 {Q^2 w \over V_{\mathrm I}},    
 \nonumber\\
T^F_{12}&=&
 -  6
 - {V^2 + w - 2 u \over Q^2 }
+{Q^2 \over u }
 + u{  V^2 -u + 4 Q^2 \over Q^2 z_1}
+ { w Q^2+Q^2_w (Q^2+V^2) \over Q^2 z_2}
 - {Q^2_w V^2 \over z_2^2}
+ Q^2_{v2} t_y {Q^2 + V^2 \over Q^2 z_1 z_2}
- 2 {Q^2 u \over V_{\mathrm I}  }, 
\nonumber\\
T^F_{13}&=&
 -{ 4 \over Q^2}
-2 {Q^2_{v2}+Q^2_u \over Q^2 z_1}
+2 {Q^2_{v2} + Q^2_w  \over Q^2 z_2}
+2 {Q^4_{v2} -Q^2 V^2 \over Q^2 z_1 z_2}
 - 2 {Q^2_{u2} \over w z_1}
 - 2 {Q^2_{w2} \over u z_2}
 + 2 {2 t+Q^2_{v2} \over V_{\mathrm I} }
+4 {Q^2_{u2} Q^2 \over V_{\mathrm I} w}
 - 4 {Q^2_{w2} Q^2 \over u V_{\mathrm I}}
 \nonumber\\&&
 - 8 {Q^6 \over u V_{\mathrm I} w}
 - 4 {Q^4 t_y \over u w z_1 z_2},
\nonumber\\ 
T^F_{14}&=&
 - {2 \over z_1}
+{2 \over z_2}
 - 4 {Q^2_u \over z_1^2}
 - 4 {Q^2_w \over z_2^2}
+4 {Q^2 \over z_1 z_2}
 - 2 {Q^2_{u2} \over w z_1}
 - 2 {Q^2_{w2} \over u z_2}
 +4 {Q^2 \over V_{\mathrm I}}
 - 4 {Q^4 t_y \over u w z_1 z_2 },  
 \nonumber\\
T^F_{15}&=&
 - {8 \over Q^2 }
 - {4 \over w }
+{ 4 \over u }
 - 4 {Q^2 \over w^2}
 - 4 {Q^2 \over u^2}
-4 {2 Q^2_{v2}-z_2 \over Q^2 z_1}
+4 {2 Q^2_{v2}+z_1 \over Q^2 z_2}
+{Q^2_u \over z_1^2 }
+{Q^2_w \over z_2^2 }
 - 2 {Q^2-V^2 \over z_1 z_2 }
 + 4 {Q^2_{v2} V^2 \over Q^2 z_1 z_2}
+2{ Q^2 \over w z_2}
 \nonumber\\&&
+2 {Q^2 \over u z_1}
+2 {u - w \over V_{\mathrm I}}
+{4 Q^2 - u \over w z_1} 
- 2 Q^2{4 Q^2 - u \over V_{\mathrm I} w}
+{4 Q^2 + w \over u z_2}
+2 Q^2 {4 Q^2 + w \over u V_{\mathrm I}}
+8 {Q^6 \over u w V_{\mathrm I}}
+8 {Q^4 t_y \over u w z_1 z_2},   
\nonumber\\
T^F_{21}&=&
 - {2 Q^2 - w+u \over Q^4}
+{z_2 - Q^2 \over Q^2 w}
 - {3 Q^2 + V^2 \over Q^2 z_1}
 - 2 {z_2 u \over Q^4 z_1}
+w {Q^2-w \over Q^4 z_2}
 - {V^2 \over z_1^2}
+t_y{2 Q^2 + V^2 \over Q^2 z_1 z_2}
+ {3 z_2+2 Q^2 \over V_{\mathrm I}}
 \nonumber\\&&
+{2 Q^2_w (u^2+w^2)+3 u^2 z_2 \over 2 Q^4 V_{\mathrm I}}
 - 2 {Q^2_w w \over z_2 V_{\mathrm I}}
 - {Q^4_w w^2 \over Q^4 z_2 V_{\mathrm I}}
+3 {z_2 u^2 V^2 \over 2 Q^4 V_{\mathrm I} z_1}
 +{z_1 w^3 \over 2 Q^4 V_{\mathrm I} z_2}
-w{(Q^2_u+Q^2_w)^2  \over V_{\mathrm I}^2},   
\nonumber\\
T^F_{22}&=&
{2 Q^2 + u-w \over Q^4}
-{z_1+ Q^2 \over Q^2 u}
-u {Q^2+u \over Q^4 z_1}
 - {3 Q^2 + V^2 \over Q^2 z_2}
 - 2 {z_1 w \over Q^4 z_2}
+{V^2 \over z_2^2}
 - {Q^2_{v2} t_y \over Q^2 z_1 z_2}
+ 2 {3 z_1-2 Q^2 \over 2 V_{\mathrm I}}
 \nonumber\\&&
-{2 Q^2_u (u^2+w^2)-3 w^2 z_1 \over 2 Q^4 V_{\mathrm I}}
+2 {Q^2_u u \over z_1 V_{\mathrm I}}
-{Q^4_u u^2 \over Q^4 z_1 V_{\mathrm I}}
+{z_2 u^3 \over 2 Q^4 V_{\mathrm I} z_1}
-3 {z_1 V^2 w^2 \over 2 Q^4 V_{\mathrm I} z_2}
-u{(Q^2_u+Q^2_w)^2  \over V_{\mathrm I}^2} ,  
\nonumber\\
T^F_{23}&=&
 - {2 \over Q^4}
+{4 \over u w}
 - 2 {Q^2_{v2} \over Q^2 z_1 z_2}
+4 {Q^2_u \over Q^2 w z_1}
+4 {Q^2_w \over Q^2 u z_2}
-4 {Q^2_u+Q^2_w \over Q^2 V_{\mathrm I}}
 - 4 {Q^2_{u2} \over V_{\mathrm I} w} 
+4 {Q^2_{w2} \over u V_{\mathrm I}} 
+8 {Q^4 \over u V_{\mathrm I} w} 
+{z_2 u^2 \over Q^4 V_{\mathrm I} z_1} 
+ {z_1 w^2 \over Q^4 V_{\mathrm I} z_2} 
 \nonumber\\&&
+4 {Q^2 t_y \over u w z_1 z_2}    ,
\nonumber\\
T^F_{24}&=&
 - {4 \over u w} 
+{4 \over z_1^2}
+{4 \over z_2^2}
+{8 \over w z_1} 
+{8 \over u z_2} 
 - 4{ Q^2 \over V_{\mathrm I} w} 
+4 {Q^2 \over u V_{\mathrm I}} 
 - 2{ z_2 \over V_{\mathrm I} z_1} 
-2 {z_1 \over V_{\mathrm I} z_2} 
+2 {(Q^2_u+Q^2_w)^2 \over V_{\mathrm I}^2}
+4{ Q^2 t_y \over u w z_1 z_2}    ,
\nonumber\\
T^F_{25}&=&
 - {4 \over Q^4}
+{14 Q^2 - 3 u \over Q^4 w} 
 - {14 Q^2 + 3 w \over Q^4 u} 
 - 4 {z_2 - Q^2 \over Q^2 w^2} 
 + 4 {z_1 + Q^2 \over Q^2 u^2} 
+{7 Q^2 + 6 u \over Q^4 z_1}
- {7 Q^2 - 6 w  \over Q^4 z_2}
 - {1 \over z_1^2}
 - {1 \over z_2^2}
 - 2{3 Q^2 + 2 V^2 \over Q^2 z_1 z_2}
 \nonumber\\&&
 - {2 \over w z_2}
 - {2 \over u z_1}
+{4 \over w z_1}
+u{6 u- Q^2 \over Q^4 w z_1}
+{4 \over u z_2}
+w{Q^2 + 6 w \over Q^4 u z_2}
+4{ Q^2 \over V_{\mathrm I} w}
-u{10 Q^2 + 3 u \over Q^2 V_{\mathrm I} w}
-4{ Q^2 \over u V_{\mathrm I}}
-w{10 Q^2 - 3 w \over Q^2 u V_{\mathrm I}}
 \nonumber\\&&
 - 3 {u^2 + w^2 \over Q^4 V_{\mathrm I}}
 +{ 20 \over V_{\mathrm I}}
 + {13 (w-u)-2 V^2  \over Q^2 V_{\mathrm I}}
 - 8 {Q^4 \over u V_{\mathrm I} w}
 - 4 {z_2 u^2 \over Q^4 V_{\mathrm I} z_1}
-4 {z_1 w^2 \over Q^4 V_{\mathrm I} z_2}
+4 {(Q^2_u+Q^2_w)^2 \over V_{\mathrm I}^2}
 - 8{ Q^2 t_y \over u w z_1 z_2}  ,
 \nonumber\\
T^F_{31}&=&
-{6 \over Q^2}
- {2 V^2 + 3 w - u \over Q^4}
+{z_2 - Q^2 \over Q^2 w}
- 3 {Q^2 + 2 V^2 \over Q^2 z_1}
- 2 {u^2 + V^4 \over Q^4 z_1}
-{2 \over z_2}
+(Q^2_w+V^2){Q^2_{w2}+Q^2_{v2}  \over Q^4 z_2}
+{V^4 \over Q^4 z_2}
- {V^2 \over z_1^2}
 \nonumber\\&&
+Q^2_{v2} V^2 {Q^2 + 2 V^2 \over Q^4 z_1 z_2}
+{2 Q^2_w (u^2+w^2)+3 u^2 z_2 \over 2 Q^4 V_{\mathrm I}}
+{3 V^2 + 4 w + 3 u \over V_{\mathrm I}}
- {Q^4_w w^2 \over Q^4 z_2 V_{\mathrm I}}
+3 {z_2 u^2 V^2 \over 2 Q^4 V_{\mathrm I} z_1}
 + {z_1 w^3 \over 2 Q^4 V_{\mathrm I} z_2}
 - w{(u + w)^2 \over V_{\mathrm I}^2  },
 \nonumber\\
T^F_{32}&=&
{6 \over Q^2}
+{ 2 V^2 - 3 u + w \over Q^4}
-{z_1 + Q^2 \over Q^2 u}
- 3 {Q^2 + 2 V^2 \over Q^2 z_2}
- 2 {w^2 + V^4 \over Q^4 z_2}
-{2 \over z_1}
+(Q^2_u+V^2){Q^2_{u2}+Q^2_{v2}  \over Q^4 z_1}
+{V^4 \over Q^4 z_1}
+{ V^2 \over z_2^2}
 \nonumber\\&&
-Q^2_{v2}V^2{ Q^2 + 2 V^2 \over Q^4 z_1 z_2}
-{2 Q^2_u (u^2+w^2)-3 w^2 z_1 \over 2 Q^4 V_{\mathrm I}}
-{3 V^2 - 3 w - 4 u \over V_{\mathrm I}}
- {Q^4_u u^2 \over Q^4 z_1 V_{\mathrm I}}
-3 {z_1 w^2 V^2 \over 2 Q^4 V_{\mathrm I} z_2}
 + {z_2 u^3 \over 2 Q^4 V_{\mathrm I} z_1}
 - u{(w+u)^2 \over V_{\mathrm I}^2}   ,
 \nonumber\\
T^F_{33}&=&
{6 \over Q^4}
+{4 \over u w}
+4 {2 Q^2_u+z_2 \over Q^4 z_1}
-4 {2 Q^2_w-z_1 \over Q^4 z_2}
-{12 \over z_1 z_2}
-2 V^2 {2 Q^2_{v2} + Q^2 \over Q^4 z_1 z_2}
+4 {Q^2_u \over Q^2 w z_1}
+4 {Q^2_w \over Q^2 u z_2}
-4 {Q^2_u + Q^2_w \over Q^2 V_{\mathrm I}}
-4 {Q^2_{u2} \over V_{\mathrm I} w}
 \nonumber\\&&
+4 {Q^2_{w2} \over u V_{\mathrm I}}
+8 {Q^4 \over u V_{\mathrm I} w}
+{z_2 u^2 \over Q^4 V_{\mathrm I} z_1}
+{z_1 w^2 \over Q^4 V_{\mathrm I} z_2}
+4 {Q^2 t_y \over u w z_1 z_2}   ,
\nonumber\\
T^F_{34}&=&
 - {4 \over u w}
+{4 \over z_1^2}
+{4 \over z_2^2}
 -{ 8 \over z_1 z_2}
+{4 \over w z_1}
+{4 \over u z_2}
-{8 \over V_{\mathrm I}}
-2 {z_2 \over V_{\mathrm I} z_1}
-2 {z_1 \over V_{\mathrm I} z_2}
+2 {(w+u)^2 \over V_{\mathrm I}^2}
+4 {Q^2 t_y \over u w z_1 z_2},  
\nonumber\\
T^F_{35}&=&
{12 \over Q^4}
+{10 Q^2 - 3 u \over Q^4 w}
 - {10 Q^2 + 3 w \over Q^4 u}
 - 4 {z_2 - Q^2 \over Q^2 w^2}
 + 4 {z_1 + Q^2 \over Q^2 u^2}
 + {19 \over Q^2 z_1}
 - 2 {u - 4 V^2 \over Q^4 z_1}
 - {19 \over Q^2 z_2}
 - 2{4 V^2 + w \over Q^4 z_2}
 - {1 \over z_1^2}
 - {1 \over z_2^2}
 \nonumber\\&&
+{10 \over z_1 z_2}
-8 {Q^2_{v2} V^2 \over Q^4 z_1 z_2}
 - {2 \over w z_2}
 - {2 \over u z_1}
  - {4 \over w z_1}
 - u{Q^2 - 6 u \over Q^4 w z_1}
 - {4 \over u z_2}
 + w{Q^2 + 6 w \over Q^4 u z_2}
  +8 {Q^2 \over V_{\mathrm I} w}
  -3 u{2 Q^2+u \over Q^2 V_{\mathrm I} w}
 - 8 {Q^2 \over u V_{\mathrm I}}
 \nonumber\\&&
 - 3 w{2 Q^2 - w \over Q^2 u V_{\mathrm I}}
+{2 V^2 + 5 w - 5 u \over Q^2 V_{\mathrm I}}
- 3 {u^2 + w^2 \over Q^4 V_{\mathrm I}}
-{8 \over V_{\mathrm I}}
 - 8 {Q^4 \over u V_{\mathrm I} w}
 - 4 {z_2 u^2 \over Q^4 V_{\mathrm I} z_1}
 -4 {z_1 w^2 \over Q^4 V_{\mathrm I} z_2}
+4 {(u+ w)^2  \over V_{\mathrm I}^2}
 - 8 {Q^2 t_y \over u w z_1 z_2},   
 \nonumber\\
T^F_{41}&=&
 - {Q^2_{u2} \over Q^4}
+{z_2 - Q^2 \over Q^2 w}
+{2 \over z_1}
+{3 u + V^2 \over Q^2 z_1}
+u{u + 2 V^2 \over Q^4 z_1}
-{ w^2 \over Q^4 z_2}
+{V^2 \over z_1^2}
+w{Q^2 + 2 w \over Q^2 u z_2}
 - {3 Q^2 + 4 w \over V_{\mathrm I}}
 - 2 {u z_2 + w^2 \over Q^2 V_{\mathrm I}}
 \nonumber\\&&
 - 3 {z_2 u^2 \over 2 Q^4 V_{\mathrm I}}
+w{Q^2 + 2 w \over u V_{\mathrm I}}
 - 3 {z_2 u^2 V^2 \over 2 Q^4 V_{\mathrm I} z_1}
 + {z_1 w^3 \over 2 Q^4 V_{\mathrm I} z_2}
+w (w+u){Q^2_w+Q^2_u  \over V_{\mathrm I}^2 }, 
\nonumber\\
T^F_{42}&=&
 -{ Q^2_{w2} \over Q^4}
+{z_1 + Q^2 \over Q^2 u}
-{2 \over z_2}
+{3 w - V^2 \over Q^2 z_2}
-w{w - 2 V^2 \over Q^4 z_2}
+ {u^2 \over Q^4 z_1}
+{V^2 \over z_2^2}
-u{Q^2 - 2 u \over Q^2 w z_1}
 - {3 Q^2 - 4 u \over V_{\mathrm I}}
 - 2 {w z_1 +u^2 \over Q^2 V_{\mathrm I}}
 \nonumber\\&&
 + 3 {z_1 w^2 \over 2 Q^4 V_{\mathrm I}}
+u{Q^2 - 2 u \over w V_{\mathrm I}}
 - 3 {z_1 w^2 V^2 \over 2 Q^4 V_{\mathrm I} z_2}
 - {z_2 u^3 \over 2 Q^4 V_{\mathrm I} z_1}
+u (w+u){Q^2_w+Q^2_u  \over V_{\mathrm I}^2}  , 
\nonumber\\
T^F_{43}&=&
{2 \over Q^2 z_1}
+{2 \over Q^2 z_2}
  - 2 {u^2 - w^2 \over Q^4 V_{\mathrm I}}
 + 2 {u + w \over Q^2 V_{\mathrm I}}
-4 {Q^2_u \over V_{\mathrm I} z_1}
+2 {Q^4_u u \over Q^4 V_{\mathrm I} z_1}
 - 4{ Q^2_w \over V_{\mathrm I} z_2}
 - 2 {Q^4_w w \over Q^4 V_{\mathrm I} z_2}
 - {z_2 u^2 \over Q^4 V_{\mathrm I} z_1}
 + {z_1 w^2 \over Q^4 V_{\mathrm I} z_2},  
 \nonumber\\
T^F_{44}&=&
 - {4 \over z_1^2}
+{4 \over z_2^2}
+2 {Q^2_u+z_2 \over V_{\mathrm I} z_1}
+2 {Q^2_w-z_1 \over V_{\mathrm I} z_2}
-2(u + w) {Q^2_w+Q^2_u  \over V_{\mathrm I}^2 }, 
\nonumber\\
T^F_{45}&=&
3 {4 Q^2 - u \over Q^4 w}
+3 {4 Q^2 + w \over Q^4 u}
 - 4 {z_2- Q^2 \over Q^2 w^2}
-4 {z_1+ Q^2 \over Q^2 u^2}
 - 3 {Q^2 + 2 u \over Q^4 z_1}
 - 3 {Q^2 - 2 w \over Q^4 z_2}
+{1 \over z_1^2}
 -{ 1 \over z_2^2}
 -{ 2 \over w z_1}
 - u{9 Q^2 + 2 u \over Q^4 w z_1}
 \nonumber\\&&
 - {2 \over w z_2}
+{2 \over u z_1}                                                                            
+{2 \over u z_2}                                            
+w{2 w - 9 Q^2  \over Q^4 u z_2}                                            
-5Q^2_u {u^2 - w^2  \over Q^4 u V_{\mathrm I}}                                         
- {u + w \over Q^2 V_{\mathrm I}}                                         
+(u+w){8 Q^4 + 5 u^2  \over Q^2 V_{\mathrm I} u w}                                                   
+4 {z_2 u^2 \over Q^4 V_{\mathrm I} z_1}
-4 {z_1 w^2 \over Q^4 V_{\mathrm I} z_2}
 \nonumber\\&&
-4(u + w) { Q^2_w+Q^2_u  \over V_{\mathrm I}^2  },  
\nonumber\\                 
T^F_{51}&=&
{1 \over 2 Q^2}
+{1 \over w}
 - {3 u + V^2 \over 2 Q^2 z_1}
 - {2 \over z_1}
 - {w \over 2 Q^2 z_2}
 - {V^2 \over z_1^2}
+{Q^4_w \over Q^2 V_{\mathrm I}}
+{4 Q^2 + u \over 2 V_{\mathrm I}}
 - {Q^2_w w \over z_2 V_{\mathrm I}}
 - w^2{Q^2_w+V^2  \over 2 Q^2 V_{\mathrm I} z_2}
+u w{u + z_2  \over 2 Q^2 V_{\mathrm I} z_1},
    \nonumber\\
T^F_{52}&=&
{1 \over 2 Q^2}
 - {1 \over u}
 -{ u \over 2 Q^2 z_1}
+ {2 \over z_2}
+{V^2 - 3 w \over 2 Q^2 z_2}
 - {V^2 \over z_2^2}
 + {Q^4_u \over Q^2 V_{\mathrm I}}
 + {4 Q^2 - w \over 2 V_{\mathrm I}}
 - {Q^2_u u \over V_{\mathrm I} z_1}
 + u^2{Q^2_u+V^2 \over 2 Q^2 V_{\mathrm I} z_1}
+u w{w+z_1  \over 2 Q^2 V_{\mathrm I} z_2},
    \nonumber\\
T^F_{53}&=&
- {2 \over w z_1}
+{2 \over u z_2}
 - {u + w \over Q^2 V_{\mathrm I}}
+{u^2 \over Q^2 V_{\mathrm I} z_1}
+{w^2 \over Q^2 V_{\mathrm I} z_2},  
\nonumber\\
T^F_{54}&=&
{4 \over z_1^2}
 -{ 4 \over z_2^2}
+{4 \over w z_1}
 -{ 4 \over u z_2}
 - 2 {Q^2 \over V_{\mathrm I} w}
 - 2 {Q^2 \over u V_{\mathrm I}}
 - {z_2 \over V_{\mathrm I} z_1}
 + {z_1 \over V_{\mathrm I} z_2},  
 \nonumber\\
T^F_{55}&=&
- {4 \over Q^2 w}
 - {2 \over Q^2 u}
 - {6 \over w^2}
+{4 \over u^2}
+{2 \over Q^2 z_1}
+{2 \over Q^2 z_2}
+{1 \over z_1^2}
+{1 \over z_2^2}
 - 2 {Q^2 - 3 u \over Q^2 w z_1}
+{2 \over w z_2}
 - {2 \over u z_1}
+2 {Q^2 + 2 w \over Q^2 u z_2}
 - 2 {w \over Q^2 V_{\mathrm I}}
 \nonumber\\&&
 - 2 {2 Q^2+u \over V_{\mathrm I} w}
 - 2 {2 Q^2-w \over u V_{\mathrm I}}
 - 2u {2 u + V^2  \over Q^2 V_{\mathrm I} z_1}
-2 {z_1 w \over Q^2 V_{\mathrm I} z_2} ,
\nonumber\\
T^F_{61}&=&
{3 \over 2 Q^2}
+{1 \over w}
+{3 \over z_1}
+{u + 3 V^2 \over 2 Q^2 z_1}
 - {2 \over z_2}
 - {2 V^2 + 3 w \over 2 Q^2 z_2}
+{V^2 \over z_1^2}
 -{ Q^2_{v2} V^2 \over Q^2 z_1 z_2}
 - {2 V^2 + 2 w + 3 u  \over 2 V_{\mathrm I}} 
+w^2 {Q^2_w-V^2 \over 2 Q^2 z_2 V_{\mathrm I}} 
 - u w {2 u + V^2  \over 2 Q^2 V_{\mathrm I} z_1}, 
     \nonumber\\
T^F_{62}&=&
 - {3 \over 2 Q^2} 
+{1 \over u}
-{ 2 \over z_1}
+{3 u - 2 V^2 \over 2 Q^2 z_1} 
+{3 \over z_2}
+{3 V^2 - w \over 2 Q^2 z_2} 
 -{ V^2 \over z_2^2}
+{Q^2_{v2} V^2 \over Q^2 z_1 z_2} 
-{ 3 w + 2 u-2 V^2  \over 2 V_{\mathrm I}} 
+u^2 {Q^2_u-V^2 \over 2 Q^2 V_{\mathrm I} z_1} 
+u w{2 w-V^2  \over 2 Q^2 V_{\mathrm I} z_2},   
\nonumber\\
T^F_{63}&=&
 - {2 \over Q^2 z_1} 
+{2 \over Q^2 z_2} 
+2{ Q^2_{v2} \over Q^2 z_1 z_2} 
 - 2 {Q^2_u \over Q^2 w z_1} 
 - 2 {Q^2_w \over Q^2 u z_2} 
 - {u - w - 4 Q^2 \over Q^2 V_{\mathrm I}} 
+2 {Q^2_{u2} \over V_{\mathrm I} w} 
 - 2{ Q^2_{w2} \over u V_{\mathrm I}} 
 - 8 {Q^4 \over u V_{\mathrm I} w} 
 - {u^2 \over Q^2 V_{\mathrm I} z_1} 
 \nonumber\\&&
+{w^2 \over Q^2 V_{\mathrm I} z_2} 
 - 4 {Q^2 t_y \over u w z_1 z_2},    
 \nonumber\\
T^F_{64}&=&
 - {4 \over z_1^2}
 - {4 \over z_2^2}
 - {2 \over w z_1} 
 - {2 \over u z_2} 
+{2 \over V_{\mathrm I}}
+{z_2 \over V_{\mathrm I} z_1} 
 +{z_1 \over V_{\mathrm I} z_2} 
 - 4{ Q^2 t_y \over u w z_1 z_2} ,  
 \nonumber\\
T^F_{65}&=&
 -{ 4 \over Q^2 w} 
+{2 \over Q^2 u} 
 - {6 \over w^2}
 -{ 4 \over u^2}
 - {6 \over Q^2 z_1} 
+{6 \over Q^2 z_2} 
 - {1 \over z_1^2}
+{1 \over z_2^2}
+2{3 Q^2 + 2 V^2 \over Q^2 z_1 z_2} 
 - 2 {u \over Q^2 w z_1} 
+{2 \over w z_2} 
+{2 \over u z_1} 
+{2 (2 u - w) \over Q^2 V_{\mathrm I}} 
 \nonumber\\&&
 - 2 {2 Q^2 - 3 u \over V_{\mathrm I} w} 
+2 {Q^2_{w2} \over u V_{\mathrm I}} 
+8 {Q^4 \over u V_{\mathrm I} w} 
+2u {2 u + V^2 \over Q^2 V_{\mathrm I} z_1} 
-2 {z_1 w \over Q^2 V_{\mathrm I} z_2} 
+8 {Q^2 t_y \over u w z_1 z_2}.   
\nonumber
\end{eqnarray}

\end{widetext}

 \noindent{\bf Acknowledgments}. The authors are grateful to Harut Avakian and Volker Burkert for
interesting discussions and comments. This work was supported by DOE contract No. DE- AC05-06OR23177, under which Jefferson Science Associates, LLC operates Jefferson Lab.
 
\bibliography{exact}{}

\end{document}